\documentclass[journal=jctcce,manuscript=article,layout=traditional]{achemso}

\usepackage{comment}
\usepackage{bbm}
\usepackage[dvipsnames]{xcolor}
\usepackage{mathrsfs}
\usepackage{subcaption}
\usepackage{epsfig}
\usepackage{soul,color}
\usepackage{graphicx}
\usepackage{mleftright}
\usepackage{amsfonts}
\usepackage{amsthm}
\usepackage[figuresright]{rotating}
\usepackage{amssymb}
\usepackage{amsmath}
\usepackage{dcolumn}
\usepackage{physics}
\usepackage{float}
\usepackage{bm}
\usepackage{verbatim}
\usepackage{braket}
\usepackage{comment}
\usepackage[normalem]{ulem}
\usepackage[ruled,vlined,linesnumbered]{algorithm2e}
\usepackage{setspace}
\usepackage{stackengine}
\usepackage{qcircuit}
\usepackage{filecontents}
\usepackage{multirow}
\usepackage{diagbox}

\usepackage{fancyhdr}

\newcommand{\swap}{\mathtt{SWAP}}

\usepackage{xr-hyper}
\usepackage[colorlinks,linkcolor=blue,anchorcolor=blue,citecolor=blue,urlcolor=blue]{hyperref}
\makeatletter

\newcommand*{\addFileDependency}[1]{
\typeout{(#1)}
\@addtofilelist{#1}
\IfFileExists{#1}{}{\typeout{No file #1.}}
}\makeatother

\newcommand*{\myexternaldocument}[1]{%
\externaldocument{#1}%
\addFileDependency{#1.tex}%
\addFileDependency{#1.aux}%
}
\myexternaldocument{supplement}

\definecolor{lightgray}{rgb}{0.83, 0.83, 0.83}

\newcommand{\innp}[1]{\langle #1\rangle}

\title{Bootstrap Embedding on a Quantum Computer}

\author{Yuan Liu}
 \email{yuanliu@mit.edu}
 \affiliation{Department of Physics, Co-Design Center for Quantum Advantage, Massachusetts Institute of Technology, Cambridge, Massachusetts 02139, USA}
\author{Oinam R. Meitei}
 \affiliation{Department of Chemistry, Massachusetts Institute of Technology, Cambridge, Massachusetts 02139, USA}
\author{Zachary E. Chin}
 \affiliation{Department of Physics, Co-Design Center for Quantum Advantage, Massachusetts Institute of Technology, Cambridge, Massachusetts 02139, USA}
 \author{Arkopal Dutt}
 \affiliation{Department of Mechanical Engineering, Massachusetts Institute of Technology, Cambridge, Massachusetts 02139, USA}
\author{Max Tao}
 \affiliation{Department of Physics, Co-Design Center for Quantum Advantage, Massachusetts Institute of Technology, Cambridge, Massachusetts 02139, USA}
\author{Isaac L. Chuang}
 \affiliation{Department of Physics, Co-Design Center for Quantum Advantage, Massachusetts Institute of Technology, Cambridge, Massachusetts 02139, USA}
 \altaffiliation{Department of Electrical Engineering and Computer Science, Massachusetts Institute of Technology, Cambridge, Massachusetts 02139, USA}
 \author{Troy Van Voorhis}
 \email{tvan@mit.edu}
 \affiliation{Department of Chemistry, Massachusetts Institute of Technology, Cambridge, Massachusetts 02139, USA}

\date{\today}

\begin{document}







\begin{abstract}
We extend molecular bootstrap embedding to make it appropriate for implementation on a quantum computer.  This enables solution of the electronic structure problem of a large molecule as an optimization problem for a composite Lagrangian governing fragments of the total system, in such a way that fragment solutions can harness the capabilities of quantum computers.  By employing state-of-art quantum subroutines including the quantum $\swap$ test and quantum amplitude amplification, we show how a quadratic speedup can be obtained over the classical algorithm, in principle.
Utilization of quantum computation also allows the algorithm to match -- at little additional computational cost -- full density matrices at fragment boundaries, instead of being limited to 1-RDMs.
Current quantum computers are small, but quantum bootstrap embedding provides a potentially generalizable strategy for harnessing such small machines through quantum fragment matching.

\end{abstract}

\maketitle

\section{Introduction}
\label{sec:intro}

Determining the ground state of large-scale
interacting fermionic systems is an important challenge in quantum chemistry, materials science, and condensed matter physics.
Just as electronic properties of molecules underpin their chemical reactivity \cite{fukui1952molecular,parr1984density,greeley2002electronic}, phase diagrams of solid state materials are also determined to a large degree by their ground state electronic structure \cite{leblanc2015solutions,zheng2015computation,kotliar2006electronic}. However, exact solution to the time-independent Schrodinger equation of a practical many-electron system remains a daunting task because the dimension of the underlying Hilbert space grows exponentially with the number of orbitals, and the computational resources required to perform calculations over such a large space can quickly exceed the capacity of current classical or quantum hardware.

One promising approach to fit a large electronic structure problem into a limited amount of computational resources is to break the original system into smaller fragments, where each fragment can be solved individually from which a solution to the whole is then obtained \cite{gordon2012fragmentation,jones2020embedding,sun2016quantum}. Efforts along this direction have successfully led to various embedding schemes that significantly expand the complexity of the systems solvable using classical computational resources, such as density-based embedding theories \cite{wesolowski2015frozen,libisch2014embedded}, density-matrix embedding theories (DMET) \cite{knizia2012density,knizia2013density,wouters2016practical,wouters2017five,faulstich2022pure}, various Green's function embedding theories \cite{hettler2000dynamical,ma2021quantum,lan2017generalized,rusakov2018self,biermann2003first,kotliar2006electronic} and the bootstrap embedding theory \cite{welborn2016bootstrap,ye2019bootstrap,ye2021accurate}. The essence of such embedding-based methods is to add an additional external potential to each fragment Hamiltonian and then iteratively update the potential until some conditions on certain observables of the system are matched. 
Nevertheless, due to the significant cost in solving the fragment Hamiltonian itself as the fragment size increases, the applicability of such methods are limited to relatively small fragments, which may lead to incorrect predictions in systems with long-range correlations \cite{zheng2017stripe}.
While approximate fragment solvers such as the coupled-cluster theory or many-body perturbation theory have greatly enhanced the applicability of such embedding methods at a reduced cost \cite{zhu2019coupled,shee2019coupled,lau2021regional}, these approximations tend to fail for strongly correlated systems due to limited treatment of electron correlation. In addition, because of limitations on computing $k$-electron reduced density matrices ($k$-RDMs for $k>2$), embedding and observable calculations beyond 2-RDM are difficult in general.

Quantum computers are believed to be promising in tackling electronic structure problems more efficiently \cite{bauer2020quantum}, despite evidence for an exponential advantage across chemical space has yet to be found \cite{lee2022there}.
One natural idea to circumvent the problems of classical eigensolvers is to use a quantum 
computer to treat the fragments.
By mapping each orbital to a constant (small) number of qubits, the exponentially
large (in the number of orbitals) Hilbert space of an interacting fermionic system can be encoded in only a polynomial 
number of qubits and terms. Indeed, quantum eigensolvers such as the quantum phase estimation (QPE) \cite{abrams1999quantum} algorithm has been proposed to achieve an exponential advantage given a properly prepared input state \cite{aspuru2005simulated} with non-exponentially small overlap with the exact ground state. More recently, various variants of the variational quantum eigensolver (VQE) \cite{Peruzzo2014variational,tilly2022variational,wang2019accelerated,grimsley2019adaptive,grimsley2022adaptvqe} have been demonstrated experimentally on NISQ devices
to achieve impressive performance as compared to classical methods. Moreover, $k$-RDMs (for any $k$) can be measured through quantum eigensolvers \cite{cramer2010efficient,zhao2021fermionic} that may circumvent the difficulty encountered on classical computers. Current quantum computers are small, but quantum embedding provides a way to tailor fragment sizes to fit into such small quantum machines to achieve a solution to the entire problem.

To take the full advantage of these quantum eigensolvers within the embedding framework \cite{otten2022localized,vorwerk2022quantum,mineh2022prb,li2022toward,ma2020quantum,ma2021quantum}, two open questions immediately arise as a result of the intrinsic nature of quantum systems.
Firstly, the wave function of a quantum system collapses when measured. This means any measurement of the fragment wave function is but a statistical sample (akin to Monte Carlo methods), and many measurements are needed to obtain statistical averages with sufficiently low uncertainty in order to achieve a good matching condition for the embedding.
Secondly, the best way to perform matching between fragments using results from quantum eigensolvers is not clear, 
and most likely a new approach needs to be formulated to match fragments. Admittedly, it would be straightforward to first estimate the density matrices by collecting a number of quantum samples and then use the estimated density matrices to minimize the cost function as in classical embedding theories \cite{welborn2016bootstrap,knizia2012density}. But this approach would be very costly especially given the increasing number of elements in qubit reduced density matrices (RDMs) that need to be estimated \cite{google2020hartree}.
Could there be a quantum method for matching, as opposed to a statistical sampling-based classical approach?

We address the two challenges by providing a quantum coherent matching algorithm and an adaptive sampling schedule, leading to a quantum bootstrap embedding (QBE) method based on classical bootstrap embedding \cite{welborn2016bootstrap}. Instead of matching the RDM element-by-element, the quantum matching algorithm employs a $\mathtt{SWAP}$ test \cite{barenco1997stabilization,buhrman2001quantum} to match the full RDM between overlapping regions of the fragments in parallel. Moreover, the quantum amplitude estimation algorithm \cite{brassard2002quantum,martyn2021grand} allows an extra quadratic speedup to reach a target accuracy on estimating the fragment overlap. In addition, the adaptive sampling changes the number of samples as the optimization proceeds in order to achieve an increasingly better matching conditions. 

It is important to compare the cost of quantum algorithms to classical algorithms carefully to claim any quantum advantage \cite{lee2022there}. Toward this end, to highlight the advantage of the new QBE algorithms, we systematically compare their cost against classical BE algorithms with a biased stochastic eigensolver (the variational Monte Carlo, VMC) and the exact solver (full configuration interaction, FCI) as baselines. Different versions of the QBE algorithms using either QPE or VQE as eigensolvers with classical or quantum (coherent) matching algorithms are also compared among themselves for clarity.

The present work invites a viewpoint of treating quantum computers as \emph{coherent sampling} machines which have three major advantages, as compared to their classical counterparts. First, the exponentially large Hilbert space provided by a quantum computer allows more efficient exact ground state solver (QPE) than their classical counterpart (exact diagonalization). Second, in the case of truncation for seeking approximate solutions, the abundant Hilbert space of quantum computers enable more flexible and expressive variational ansatz than classical computers, leading to more accurate solutions. Third, the coherent nature of quantum computers allows sampling to be performed at a later stage, e.g. after quantum amplitude amplification of matching conditions to extract just the feedback desired, instead of having to read out full state of a system.

The rest of the paper is organized as follows. Sec. \ref{sec:classical-be} overviews bootstrap embedding method at a high level and analyzes its scaling on classical computers, in order to motivate the need for bootstrap embedding on quantum computers. This section serves to set the notation and baseline of comparison for the rest of the paper. Sec. \ref{sec:qbe-methods} presents the theoretical framework of quantum bootstrap embedding in detail as constraint optimization problems. In Sec. \ref{sec:qbe-algorithms}, we give details of the QBE algorithm to solve the optimization problem. In Sec. \ref{sec:results}, we apply our methods to hydrogen chains under minimal basis where both classical and quantum simulation results are shown to demonstrate the convergence and sampling advantage of our QBE method. We conclude the paper in Sec. \ref{sec:conclusion} with a summary of comparisons between classical and quantum BE
discussed in the paper, as well as prospects and future directions.

\section{Ideas of Bootstrap Embedding}
\label{sec:classical-be}

The idea of Bootstrap Embedding (BE) for quantum chemistry has recently led to a promising path to tackle large-scale electronic structure problems \cite{welborn2016bootstrap,ye2019bootstrap,ye2020bootstrap}. In this section, we establish the terminology and framework that will be used in the rest of the paper. We first briefly review BE and outline the main framework of BE for computation on a classical computer in Sec. \ref{sec:embedding-hamiltonian} and \ref{sec:classical-matching-optimization} for non-chemistry readers, to set up the notation. We then begin presenting new material by discussing typical behavior and computational resource requirements for BE on classical computers in Sec. \ref{sec:classical-resource-req-behavior}, which leads to the quest for performing BE on a quantum computer in Sec. \ref{sec:quest-for-quantum-be}. 

\subsection{Fragmentation and Embedding Hamiltonians} 
\label{sec:embedding-hamiltonian}

To provide a foundation for a more concrete exposition of the bootstrap embedding method, we first establish some rigorous notation for discussing molecular Hamiltonians and their associated Hilbert spaces. We will work with the molecular Hamiltonian under the second quantization formalism. Specifically, given a particular molecule of interest, define $O = \{\phi_\mu\ |\ \mu = 1,\ldots, N\}$ to be an orthonormal set of single-particle local orbitals (LOs), where $N$ is the total number of orbitals; in this work, these LOs are generated through L\"owdin's symmetric orthogonalization method \cite{lowdin1950non}. The full Hilbert space $\mathcal{H}$ for the entire molecular system is thus given by $\mathcal{H} = \mathcal{F}(O)$, where $\mathcal{F}(O)$ denotes the Fock space determined by the LOs in the set $O$. Further define the creation (annihilation) operator $c^\dagger_\mu$ ($c_\mu$) which creates (annihilates) an electron in the LO $\phi_\mu$, the molecular Hamiltonian is written in the second-quantized notation
\begin{align}
\hat{H} = \sum_{\mu\nu=1}^N h_{\mu\nu}c_\mu^\dagger c_\nu + \frac{1}{2}\sum^N_{\mu\nu\lambda\sigma=1}V_{\mu\nu\lambda\sigma}c^\dagger_\mu c^\dagger_\nu c_\sigma c_\lambda
\label{total-ham}
\end{align}
where $h_{\mu\nu}$ and $V_{\mu\nu\lambda\sigma}$ are the standard one- and two-electron integrals.

Note that the number of terms in the full molecular Hamiltonian $\hat{H}$ scales polynomially with the total number of orbitals $N$, but the dimension of $\mathcal{H}$ scales exponentially with $N$. Clearly, for large $N$, it will become prohibitively expensive to directly compute the exact full ground state. To circumvent this issue, we divide the full molecule into multiple smaller fragments, each equipped with its own ``embedding Hamiltonian'' which contains a number of terms that only scales polynomially with the number of orbitals \textit{in the fragment}. Given that there are potentially far fewer orbitals in each fragment than in the whole molecular system, computing the ground state of each fragment's embedding Hamiltonian can be significantly less expensive than computing the ground state of the full system. Furthermore, using the bootstrap embedding procedure to be described later, the ground states of individual fragments can, to a high degree of accuracy, be algorithmically combined to recover the desired electron densities prescribed by the exact ground state of the full system. Thus, this combination of fragmentation and bootstrap embedding can be used to reconstruct the full molecular ground state more efficiently than by direct computation alone.

We now briefly review the construction of embedding Hamiltonians for each fragment. Consider a single fragment associated with a label $A$, without loss of generality, define $O^{(A)} = \{\phi_\mu\ |\ \mu = 1,\ldots, N_A\}$ with $N_A \leq N$ to be the set of LOs contained in fragment $A$; we will refer to $O^{(A)}$ as the set of fragment orbitals. Note that $O^{(A)} \subseteq O$, the set of LOs for the entire molecular system. The construction of the embedding Hamiltonian $\hat{H}^{(A)}$ for fragment $A$ begins with any solution of the ground state of the full system $\hat{H}$. For simplicity, the Hartree-Fock (HF) solution $\ket{\Phi_{\textrm{HF}}}$ is often used because it is easy to obtain on a classical computer. By invoking a Schmidt decomposition, we can write $\ket{\Phi_{\textrm{HF}}}$ with the following tensor product structure for $\forall~A$
\begin{align}\label{eq:schmidt}
    \ket{\Phi_{\textrm{HF}}} = \left(\sum_{i = 1}^{N_A} \lambda_i^{(A)}\ket{f_i^{(A)}} \otimes \ket{b_i^{(A)}} \right) \otimes \ket{\Psi_\textrm{env}^{(A)}}.
\end{align}
In the above decomposition, the $\ket{f_i^{(A)}}$ represent single-particle fragment states contained in the Fock space $\mathcal{F}(O^{(A)})$ of fragment orbitals. On the other hand, the $\ket{b_i^{(A)}}$ and $\ket{\Psi_\textrm{env}^{(A)}}$ represent Slater determinants
contained in the ``environment'' Fock space $\mathcal{F}(O\setminus O^{(A)})$ of the $N-N_A$ orbitals not included in the fragment. The key difference between the single environment state $\ket{\Psi_\textrm{env}^{(A)}}$ and the various ``bath'' states $\ket{b^{(A)}_i}$ is that the bath states $\ket{b^{(A)}_i}$ are entangled with the fragment states $\ket{f_i^{(A)}}$ while $\ket{\Psi_\textrm{env}^{(A)}}$ is not; this entanglement is quantified by the Schmidt coefficients $\lambda_i^{(A)}$. 
Crucially, since the HF solution is used, the sum in Eq. \eqref{eq:schmidt} only has $N_A$ terms (as opposed to $2^{N_A}$ for a general many-body wave function). Denote the collection of the $N_A$ entangled bath orbitals as $O^{(A)}_{\rm bath} = \{\beta_\mu\ |\mu = 1,\ldots, N_A\}$, where each of the LOs $\beta_\mu$ are linear combinations of the original LOs not included in the fragment, $\beta_\mu \in \operatorname{Span}\{O\setminus O^{(A)}\}$. Furthermore, we denote the Fock space that corresponds to this set of entangled bath orbitals as $\mathcal{F}(O^{(A)}_{\rm bath})$. 

This tensor product structure of $\ket{\Phi_{\rm HF}}$ allows us to naturally decompose the Hilbert space $\mathcal{H}$ for the full molecular system into the direct product of two smaller Hilbert spaces, namely 
\begin{align}
    \mathcal{H} = \mathcal{H}^{(A)} \otimes \mathcal{H}^{(A)}_\textrm{env},
\end{align}
where 
\begin{align}
    \mathcal{H}^{(A)} = \mathcal{F}(O^{(A)}) \otimes \mathcal{F}(O^{(A)}_{\rm bath})
\end{align}
is the active fragment embedding space and $\mathcal{H}^{(A)}_\textrm{env}$ contains the remaining states, including $\ket{\Psi_\textrm{env}^{(A)}}$. Note that since both sets $O^{(A)}$ and $O^{(A)}_{\rm bath}$ have size $ N_A$, the fragment Hilbert space $\mathcal{H}^{(A)}$ is a Fock space spanned of just $2N_A$ single-particle orbitals. The core intuition motivating this decomposition is that, in the exact ground state of the full system, states in $\mathcal{H}^{(A)}_\textrm{env}$ are unlikely to be strongly entangled with the many-body fragment states (consider the approximate HF ground state in Eq. \eqref{eq:schmidt}, where they are perfectly disentangled); therefore, in a mean-field approximation, it is reasonable to entirely disregard the states in $\mathcal{H}^{(A)}_\textrm{env}$ when calculating the ground state electron densities on fragment $A$. Following this logic, we can define an embedding Hamiltonian $\hat{H}^{(A)}$ for fragment $A$ \textit{only} on the $2N_A$ LOs in $\mathcal{H}^{(A)}$, which will have the form
\begin{align}
    \hat{H}^{(A)} = \sum^{2N_A}_{pq} h^{(A)}_{pq} a^{(A)\dagger}_p a^{(A)}_q + \frac{1}{2}\sum^{2N_A}_{pqrs} V^{(A)}_{pqrs} a^{(A)\dagger}_p a^{(A)\dagger}_q a^{(A)}_s a^{(A)}_r,
    \label{embedding-ham}
\end{align}
given some creation and annihilation operators $a^{(A)\dagger}_p$ and $a^{(A)}_p$, which respectively create and annihilate electrons in orbitals from the combined set $O^{(A)} \cup O^{(A)}_{\rm bath}$ for $\mathcal{H}^{(A)}$. The new one- and two- electron integrals $h^{(A)}_{pq}$ and $V^{(A)}_{pqrs}$ can be computed by projecting $\hat{H}$ into the smaller Hilbert space $\mathcal{H}^{(A)}$ (consult the Supporting Information (SI) Sec. \ref{app:frag} for details on constructing $h_{pq}^{(A)}$ and $V_{qprs}^{(A)}$). Note that since we can choose $2N_A \ll N$, the ground state of this embedding Hamiltonian can be solved at a significantly reduced cost when compared to that of the full system Hamiltonian.

We are hence prepared to generate an embedding Hamiltonian for any arbitrary fragment of the original molecular system. However, the ground state electron densities of the fragment embedding Hamiltonian are unlikely to exactly match those of the full system Hamiltonian because, as mentioned above, the embedding process may neglect some small (but nonzero) entanglement of the fragment orbitals with the environment. Because we can expect interactions in the molecular Hamiltonian to be reasonably local, we anticipate that the electron densities on orbitals near the edge of the fragment (those closest to the ``environment'') will deviate most significantly from their true values, while electron densities on orbitals toward the center of the fragment will be most accurate. Note that the uneven distribution of entanglement in molecular systems may likely lead to the potential sensitivity of the BE results to particular choices and partitions of fragments \cite{claudino2019automatic,waldrop2021projector,ye2019atom,knizia2013intrinsic,lau2021regional,nusspickel2022systematic}, while how quantum computers may help to reduce such dependence is an open problem.

\begin{figure*}[htbp]
\centering
\includegraphics[width=17cm]{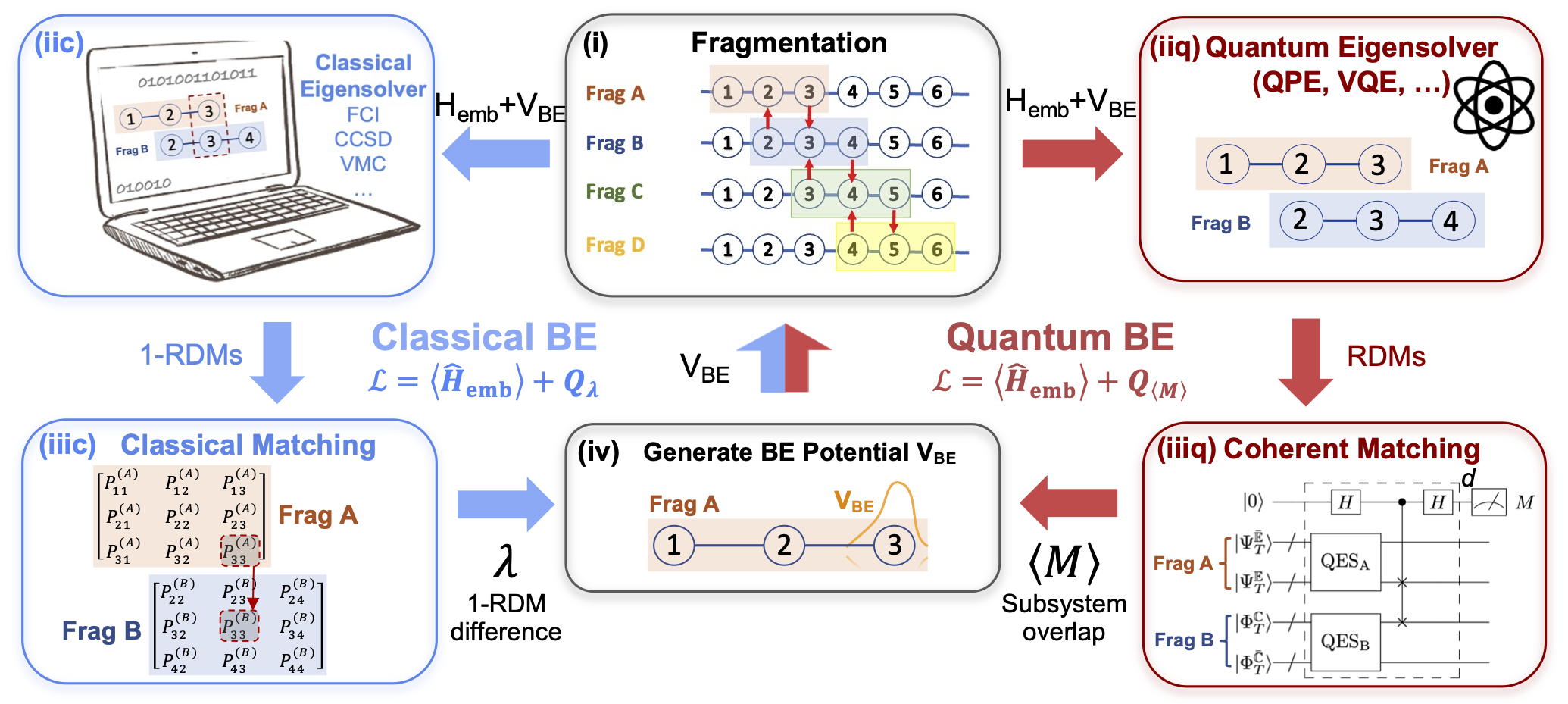}
\caption{Schematic of bootstrap embedding on classical (left, blue arrows) and quantum (right, red arrows) computers. The arrows indicate BE iterative loops that are used to optimize the corresponding objective functions. Starting from panel (i) (upper center), the original system is first broken into overlapping fragments (Fragmentation), where each fragment is solved using a classical (iic) (upper left) or quantum eigensolver (iiq) (upper right). In classical matching, the 1-electron reduced density matrices (1-RDM) on the overlapping sites of adjacent fragments are used to obtain the matching condition (iiic) (lower left), while in the quantum case a coherent matching protocol based on $\mathtt{SWAP}$ tests of overlapping sites combined with a single qubit measurement (iiiq) (lower right). The matching results are then used by classical computers to generate the bootstrap embedding potential $V_{\rm BE}$ (iv) (lower center) and the updated fragment embedding Hamiltonian $H_{\rm emb} + V_{\rm BE}$ (back to panel (i) in order to minimize a target objective function $\mathcal{L}$ in both classical and quantum case.
\label{fig:be-schematic}
}
\end{figure*}

To improve the accuracy of the fragment ground state wave function near the fragment edge, we employ the technique of bootstrap embedding. Broadly speaking, we first divide the full molecule into overlapping fragments such that the edge of each fragment overlaps with the center of another. Fig. \ref{fig:be-schematic}i illustrates this fragmentation strategy: for example, we see that the edge of fragment $A$ (labeled as orbital 3) coincides with the center of fragment $B$. We then apply additional local potentials to the edge sites of each fragment to match their electron densities to those on overlapping center sites of adjacent fragments. Because we expect the electron densities computed on the center sites to be closer to their true values, these added local potentials should improve the accuracy of each fragment wave function near the edges. In the next section, we will formalize this edge-to-center matching process rigorously and discuss its implementation on a classical computer. 

\subsection{Matching Electron Densities: an Optimization Problem}
\label{sec:classical-matching-optimization}

As mentioned in the previous section, we intend to correct the electron density error near a fragment's edge by applying a local potential to the edge; this local potential serves to match the edge electron density of the fragment to the center electron density of an adjacent overlapping fragment, which we expect to be more accurate. In principle, to achieve an exact density matching, all $k$-electron reduced density matrices ($k$-RDM, for any $k$) on the overlapping region have to be matched. However, in practice, such matching beyond the 2-RDM is difficult on a classical computer due to the mathematical challenge that the number of terms in $k$-RDM in general increases exponentially as $k$. In addition, almost all electronic structure codes available on classical computers are programmed to deal with only 1- and 2-RDMs, despite the importance of $k$-RDMs ($k > 2$) for computing observables such as entropy and other multi-point correlation functions \cite{toldin2018entanglement}. Due to this reason, the discussion of density matching process in classical BE in this section will be based on 1-RDMs. We note that the matching process applies similarly if $k$-RDMs are matched.

We begin by introducing some rigorous notation. Recall that a fragment $A$ is defined by a set of local orbitals $O^{(A)}$ which constitute the fragment. We partition this set of LOs into a subset of edge sites (or orbitals), denoted $\mathbb{E}^{(A)}$, and a subset of center sites, denoted $\mathbb{C}^{(A)}$, such that $\mathbb{E}^{(A)} \cup \mathbb{C}^{(A)} = O^{(A)}$ and $\mathbb{E}^{(A)} \cap \mathbb{C}^{(A)} = \emptyset$. Given the ground state wave function $\ket{\Psi^{(A)}}$ of the embedding Hamiltonian, we further define the 1-electron reduced density matrix (1-RDM) $\mathbf{P}^{(A)}$ according to \begin{align}
    P^{(A)}_{pq} = \bra{\Psi^{(A)}}a^{(A)\dagger}_pa^{(A)}_q\ket{\Psi^{(A)}}
    \label{1-rdm-def}
\end{align} 
where $p,q = 1,\ldots,2N_A$ and the operators $a^{(A)\dagger}_p$ and $a^{(A)}_q$ are defined in the previous section.

Suppose, for example, that the edge of fragment $A$ overlaps with the center of another fragment $B$ so that $\mathbb{E}^{(A)} \cap \mathbb{C}^{(B)} \neq \emptyset$. On a high level, the goal of bootstrap embedding is to find a ground state wave function $\ket{\Psi^{(A)}}$, perturbed by local potentials on the edge sites of $A$, such that $|P^{(A)}_{pq} - P^{(B)}_{pq}|\rightarrow 0$ for indices $p$ and $q$ that correspond to orbitals in the set of overlapping sites $\mathbb{E}^{(A)} \cap \mathbb{C}^{(B)}$. More generally, and more rigorously, the goal is to find a wave function which minimizes the fragment Hamiltonian energy
\begin{align}
    \ket{\Psi^{(A)}} = {\rm arg}\min_{\Psi^{(A)}} \innp{\hat{H}^{(A)}}_A
\end{align}
subject to the constraints
\begin{align}\label{eq:1rdm-constraint}
    \innp{a^{(A)\dagger}_p a_q^{(A)}}_A - P_{pq}^{(B)} = 0
\end{align}
for \textit{all} other fragments $B$ with $\mathbb{E}^{(A)} \cap \mathbb{C}^{(B)} \neq \emptyset$ and for all $p,q$ corresponding to orbitals in $\mathbb{E}^{(A)} \cap \mathbb{C}^{(B)}$. Here, we explicitly write the expectation $\innp{\cdot}_A = \bra{\Psi^{(A)}} \cdot \ket{\Psi^{(A)}}$ in terms of $\ket{\Psi^{(A)}}$ to indicate that the optimization is over the wave function of $A$.

We can formulate this constrained optimization problem as finding the stationary solution to a Lagrangian by associating a scalar Lagrange multiplier $(\lambda^{(A)}_{B})_{pq}$ to Eq. \eqref{eq:1rdm-constraint}. Since Eq. \eqref{eq:1rdm-constraint} has to be satisfied for any $p,q$ and $B$ that overlaps with $A$, these constraint can be rewritten in a more compact vector form $\bm{\lambda^{(A)}_{B}} \cdot \bm{\mathcal{Q}_{\textbf{1-RDM}}}(\Psi^{(A)}; \mathbf{P}^{(B)})$ where the dot product conceals the implicit sum over $p,q$, and each component of the vector $\mathcal{Q}_\textrm{1-RDM}(\Psi^{(A)}; \mathbf{P}^{(B)})_{pq}$ represents the constraint associated with Lagrange multiplier $(\lambda^{(A)}_B)_{pq}$, given by the left hand side of Eq. (\ref{eq:1rdm-constraint}). With this notation, we arrive at the following Lagrangian with the constraint added as an additional term
\begin{align}
    \mathcal{L}^{(A)} =& \langle \hat{H}^{(A)} \rangle_A +\mathcal{E}^{(A)} 
    \left(\bra{\Psi^{(A)}} \Psi^{(A)}\rangle -1 \right) \nonumber \\
    +& \sum_{B} \bm{\lambda^{(A)}_{B}}
    \cdot \bm{\mathcal{Q}_{\textbf{1-RDM}}}(\Psi^{(A)}; \mathbf{P}^{(B)})
    \label{eq:be_lagrangian_a},
\end{align}
where once again the $B$ are fragments adjacent to $A$ with $\mathbb{E}^{(A)} \cap \mathbb{C}^{(B)} \neq \emptyset$ and $p,q$ are indices of orbitals contained in the overlapping set $\mathbb{E}^{(A)} \cap \mathbb{C}^{(B)}$. 
Here, the additional constraint with Lagrange multiplier $\mathcal{E}^{(A)}$ is also included to ensure normalization of the ground state wave function $\ket{\Psi^{(A)}}$. 
Solving for the stationary solution of the Lagrangian in Eq. (\ref{eq:be_lagrangian_a}) will only result in a ground state wave function for fragment $A$ whose 1-RDM elements at the edge sites match those at the center sites of adjacent overlapping fragments. However, we would instead like to solve for such a ground state for \textit{all} fragments in the molecule simultaneously. Toward this regard,
we can combine all individual fragment Lagrangians (of the form of Eq. (\ref{eq:be_lagrangian_a})) into a single composite Lagrangian for the whole molecule, given by
\begin{align}
    \mathcal{L} = \sum_{A=1}^{N_{\rm frag}}\mathcal{L}^{(A)} + \mu \mathcal{P}
     \label{eq:be_lagrangian}
\end{align}
where $N_{\textrm{frag}}$ is the number of fragments in the molecule. Observe that we have added one additional constraint
\begin{align}
     \mathcal{P} = \left(\sum_{A=1}^{N_{\rm frag}}\sum_{p' \in \mathbb{C}^{(A)}} \innp{a^{(A)\dagger}_{p'} a_{p'}^{(A)}}_A\right) - N_e
     \label{global-constraint}
\end{align}
with Lagrange multiplier $\mu$ to restore the desired total number of electrons in the molecule, $N_e$. Note in Eq. (\ref{global-constraint}) that $p'$ is summed over indices corresponding to orbitals only in $\mathbb{C}^{(A)}$; this is to ensure that there is no double-counting of electrons in the whole molecule. By self-consistently finding ground states $\ket{\Psi^{(A)}}$ for $A = 1,\ldots,N_{\textrm{frag}}$ which make the composite Lagrangian in Eq. (\ref{eq:be_lagrangian}) stationary, we will have completed the density matching procedure for all fragments, and the process of bootstrap embedding will be complete. 

We can gain insight into which wave functions $\ket{\Psi^{(A)}}$ will make the composite Lagrangian $\mathcal{L}$ stationary by differentiating $\mathcal{L}$ with respect to $\ket{\Psi^{(A)}}$ for some fixed fragment $A$ and setting the resulting expression equal to zero. Upon some algebraic manipulation, we can recover the eigenvalue equation
\begin{align}
    ( \hat{H}^{(A)} + V_{\rm BE} ) \ket{\Psi^{(A)}} = -\mathcal{E}^{(A)} \ket{\Psi^{(A)}},
    \label{eig-equation}
\end{align}
where $V_{\textrm{BE}}$, the local bootstrap embedding potential, is given by
\begin{align}\label{eq:VBE}
    V_{\rm BE} = \sum_B\sum_{p,q} (\lambda_B^{(A)})_{pq} a^{(A)\dagger}_p a^{(A)}_q + \mu \sum_{p'} a^{(A)\dagger}_{p'} a_{p'}^{(A)} 
\end{align} 
where the $p,q$ are indices of orbitals in the overlapping set $\mathbb{E}^{(A)} \cap \mathbb{C}^{(B)}$, and the $p'$ are indices of orbitals in the fragment center $\mathbb{C}^{(A)}$. We see that, when the composite Lagrangian is made stationary with respect to the fragment wave functions, the bare fragment embedding Hamiltonians become dressed with a potential $V_{\textrm{BE}}$ that contains a component local to the edge sites of each fragment (see the left term of Eq. (\ref{eq:VBE})). This observation confirms our intuition that adding a local potential to the edge of one fragment will allow the edge site electron density to be matched to that of a center site on an overlapping neighbor. Note that $V_{\rm BE}$ also contains an additional potential on the center sites of each fragment (see the right term of Eq. (\ref{eq:VBE})); this is simply to conserve the total electron number in the molecule. Moreover, $V_{\rm BE}$ as in Eq. \eqref{eq:VBE} only contains one-body terms because only 1-RDM is used for density matching. In general, $V_{\rm BE}$ will contain up to $k$-body terms if $k$-RDMs are used for matching.

On a classical computer, the composite Lagrangian in Eq. (\ref{eq:be_lagrangian}) is made stationary through an iterative optimization algorithm \cite{welborn2016bootstrap} until the edge-to-center matching condition for all fragments is satisfied by some criterion. One possible criterion is to terminate the algorithm when the root-mean-squared 1-RDM mismatch, given by 
\begin{align}
    \epsilon = \left[ \frac{1}{N_{\rm sites}} \sum_A^{N_{\rm frag}}\sum_{B}\sum_{p,q}(P^{(A)}_{pq} - P^{(B)}_{pq})^2\right]^{\frac{1}{2}},
    \label{1rdm-mismatch}
\end{align}
drops below some predetermined threshold. Note again that $p,q$ are indices corresponding to orbitals in the overlapping set $\mathbb{E}^{(A)} \cap \mathbb{C}^{(B)}$; also, $N_{\rm sites}$ denotes the total number of overlapping sites in the whole molecule, equal to $N_{\rm sites} =  \sum_A^{N_{\rm frag}}\sum_{B}\sum_{p,q} 1$.
The final set of density-matched fragment wave functions $\{\ket{\Psi^{(A)}}\}$ for $A = 1,\ldots,N_{\rm frag}$ which solve the composite Lagrangian can then be used to reconstruct the electron densities and other observables for the full molecular system, as desired. 

\subsection{Resource Requirement and Typical Behavior of BE on Classical Computers}
\label{sec:classical-resource-req-behavior}

Given the notation established for classical BE, we now begin presenting new material. We discuss the computational resource requirement and typical behaviors of performing BE on classical computers to set the stage for a quantum BE theory. The details of the classical BE algorithms are omitted for succinctness, and we refer the reader to Ref. \cite{welborn2016bootstrap,ye2019bootstrap,ye2020bootstrap,ye2021accurate} for details.

\begin{figure}[ht!]
    \centering
    \includegraphics[width=9cm]{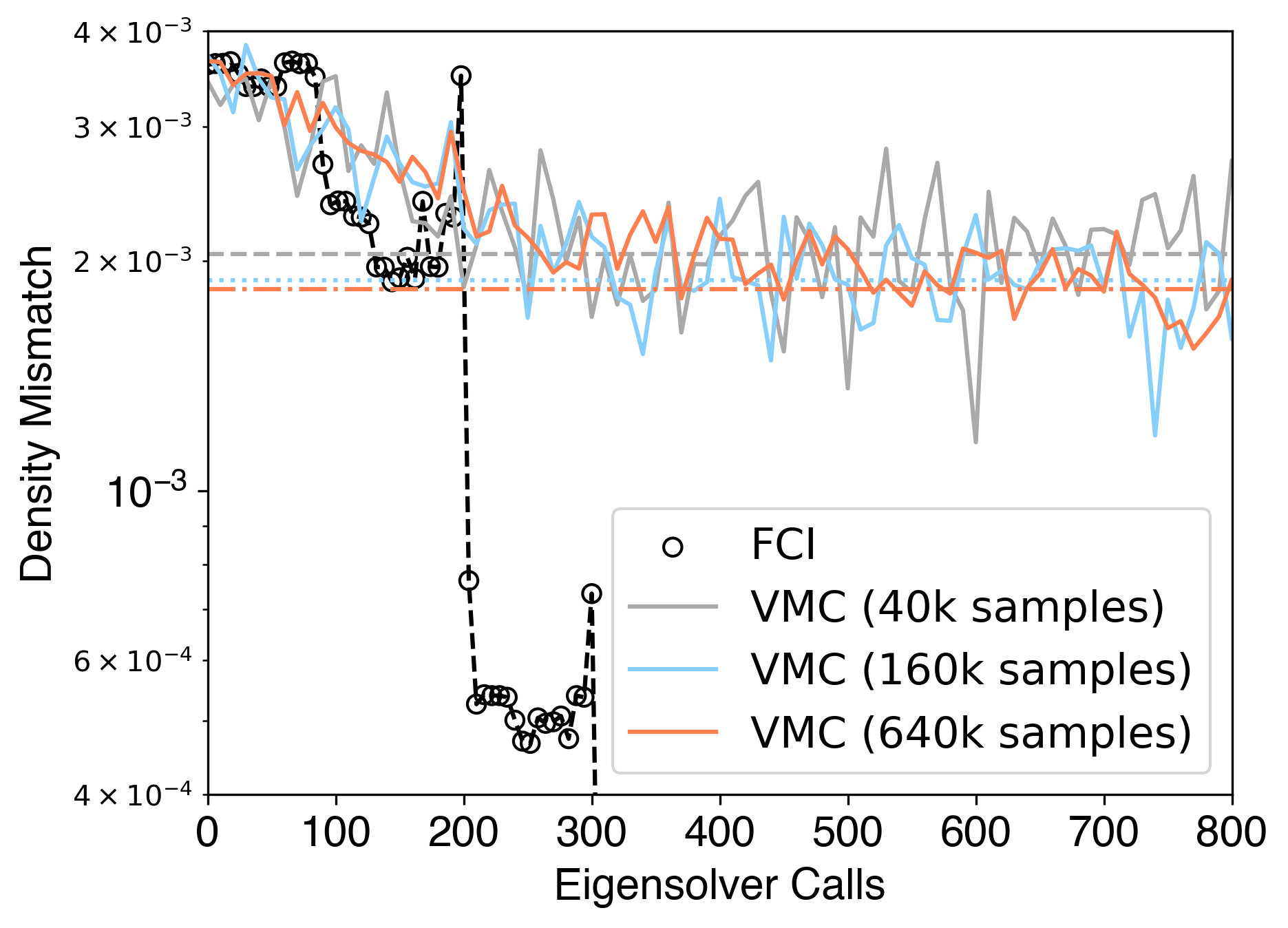}
    \caption{\label{fig:err-iter_classical} Typical convergence of density mismatch with respect to the number of eigensolver calls in classical bootstrap embedding with a deterministic eigensolver (FCI, black circle)  and a stochastic eigensolver (VMC) with different number of samples (grey, blue, and orange solid lines). The horizontal dashed lines shows the final plateaued value of the density mismatch for VMC, while the FCI data converges to $10^{-6}$ after 700 eigensolver calls (not shown on the figure). The discrete jumps around 200 and 300 eigensolver calls are due to switching to the next BE iteration. The data is obtained for an H$_8$ linear chain under STO-3G basis. See SI Sec. \ref{app:be-vmc} for computational details.
    }
    \label{fig:classical_baseline}
\end{figure}

The space and time resource requirement to perform the classical BE can be broken down into two parts: a) the number of iteration steps to reach a fixed accuracy for $\epsilon$ (Eq. \eqref{1rdm-mismatch}); b) the runtime of the fragment eigensolver. For a), numerical evidence suggests an exponentially fast convergence on total system energy as the number of bootstrap iteration increases (black trace in Fig. \ref{fig:classical_baseline} for FCI), while a proof of the convergence rate has yet to be established.

We focus on resource requirement in b) in the following. Admittedly, an exact classical eigensolver such as full configuration interaction (FCI) can be used to solve the embedding Hamiltonian in Eq. \eqref{embedding-ham}. However, both the storage space and time requirement scales exponentially as the the number of orbitals (see blue symbols and dashed line in Fig. \ref{fig:runtime-systemsize} for the runtime scaling of FCI). Even with the state-of-the-art classical computational resources, exact solutions using FCI are only tractable for systems up to 20 electrons in 20 orbitals \cite{vogiatzis2017pushing}.

As a result, classical computation of BE resorts to approximate eigensolvers with only polynomial cost in practice, by properly truncating or sampling from the fragment Hilbert space. One example for truncation is the coupled-cluster singles and doubles (CCSD) \cite{bartlett2007coupled}, which scales with $N^6$ with $N$ being the number of orbitals. Alternately, different flavors of stochastic electronic structure solvers
can be employed as fragment solvers in BE. Depending on implementation, these stochastic solvers can be biased
or unbiased
(if unbiased, with a cost of introducing the phase problem in general) \cite{morales2021frontiers,lee_twenty_2022,shee2019achieving,liu2018ab}. Collecting each sample on a classical computer usually has similar cost as a mean field theory (roughly $O(N^3)$), while the overall target accuracy $\epsilon$ on observable estimation can be achieved 
with a sampling overhead of roughly $O(\frac{1}{\epsilon^2})$ with a constant prefactor depending on the severity of the sign problem.

Importantly, the sampling feature of these stochastic electronic structure methods on classical computers are strikingly similar to the nature of quantum computers where measurement necessarily collapses the wave function. As a result, only a classical sample (in terms of measurement results) can be obtained from a quantum computer. This similarity suggests a general strategy that many sampling techniques in stochastic classical algorithms can be deployed to design better quantum algorithms. For example, sophisticated importance sampling techniques \cite{nightingale1998quantum,foulkes2001quantum} can be employed to greatly improve the sampling efficiency in both classical \cite{liu2018ab} and quantum cases \cite{huang2020predicting}. 

Due their shared feature on sampling between classical stochastic algorithm and quantum eigensolvers, we shall use one approximate sign-problem-free flavor of stochastic electronic structure method, the variational Monte Carlo (VMC), to serve as an additional baseline scenario for comparison with quantum BE in later sections.
In addition to BE convergence behavior with a FCI solver, Fig. \ref{fig:classical_baseline} also shows, for a VMC eigensolver with single Slater-Jastrow type wave function with two-body Jastrow factors \cite{wagner2007energetics,wheeler2021pyqmc}, the density mismatch converges exponentially fast initially as iteration number increases with varying number of samples. However, partially due to the statistical noise on estimating the 1-RDM (thus the gradient for the optimization), the final density mismatch plateaus to a finite biased value. Comparing the VMC results across different numbers of samples, we can see that the bias improves as the number of samples increases (dashed horizontal lines). It is also evident that the orange trace (640k samples) has smaller fluctuation as compared to blue (160k samples) and grey (40k samples). Strictly speaking, an increase of sample size by a factor of 16 would decrease the statistical fluctuations by a factor of 4. However, our numerical data in Fig. \ref{fig:classical_baseline} only shows \emph{qualitative} but not quantitative agreement with this statement. We attribute part of the bias in the plateau to the intrinsic truncation of the VMC ansatz in addition to statistical fluctuations.

The increasing accuracy of density mismatch with respect to BE iteration also suggests an increasing number of samples are needed.
Thus, an optimal number of samples at each BE iteration must be determined to achieve the desired accuracy in the matching conditions. A careful design of such a sampling schedule can potentially save a large amount of computational resources. We defer a thorough discussion of this point to later sections on quantum BE.

\subsection{The Quest for BE on Quantum Computers}
\label{sec:quest-for-quantum-be}

By employing the coherent superposition and entanglement of quantum states, the limitation of an exact classical solver can be overcome by substituting it with an exact quantum eigensolver such as the quantum phase estimation (QPE) algorithm \cite{abrams1999quantum}. This section directly compares the cost between the two exact eigensovlers on quantum and classical computers, the QPE and the FCI solvers, using hydrogen chains where the initial trial state with a non-vanishing overlap with the exact eigenstate for QPE can be efficiently prepared on classical computers.

\begin{figure}[ht!]
    \centering
    \includegraphics[width=9cm]{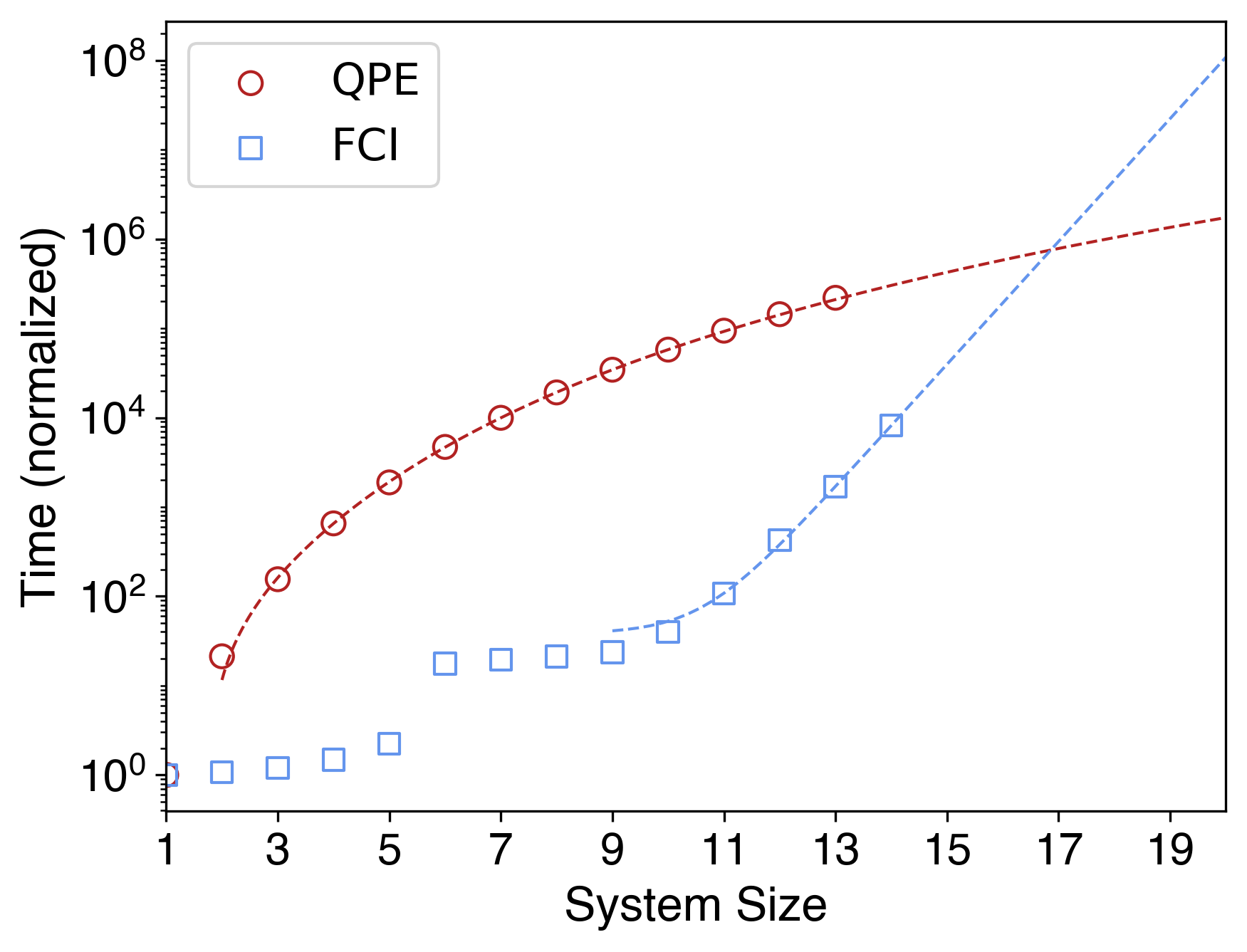}
    \caption{\label{fig:runtime-systemsize} Runtime (normalized) as a function of system size $n$ for finding the ground state of a linear hydrogen chain H$_n$ at STO-3G basis, comparing an exact classical solver (FCI, blue square) and an exact quantum solver (QPE, red circle) on real classical and quantum devices. Red (blue) dashed line shows a polynomial (exponential) fit to the QPE (FCI) runtime. Note the crossover at large system size. 
    }
\end{figure}

Fig. \ref{fig:runtime-systemsize} compares the runtime (gate depth) of FCI and QPE for finding the ground state of linear hydrogen chain H$_n$ for different system size $n$. Clearly, the QPE runtime scales only polynomially as the system size increases as expected \cite{aspuru2005simulated,lee2022there}, while its classical counterpart (FCI) has an exponentially increasing runtime. Note the runtime is normalized to the case of $n=1$ for each solver separately (see SI Sec. \ref{app:computational-details}). The dramatic advantage in the runtime scaling of quantum over classical eigensolvers demonstrated above suggests formulating BE on a quantum computer can bring significant benefits. 

One might think that the eigensolver at the heart of the classical BE algorithm could simply be replaced with a quantum one. However, as mentioned before, there are two outstanding challenges for such a quantum bootstrap embedding (QBE) method. First, just as in classical stochastic methods, the results of a quantum eigensolver need to be measured for later use, but quantum wave functions collapse after measurement. Therefore, sampling from the quantum eigensolver is required, and the optimal sampling strategy is unclear. Secondly, with quantum wave function from quantum eigensolvers, it is not wise to achieve matching between fragments in the same way as classical BE, as many incoherent samples are needed to obtain a good estimation of the 1-RDM elements. Clearly, performing matching in a quantum way is desired.

In the next two sections (Secs. \ref{sec:qbe-methods} and \ref{sec:qbe-algorithms}), we present how we address these two challenges by an adaptive quantum sampling scheduling algorithm and a quantum coherent matching algorithm in detail.

\section{Quantum Bootstrap Embedding Methods}
\label{sec:qbe-methods}

In previous sections, we have seen potential advantages of performing bootstrap embedding on quantum computers, and discussed two major challenges of doing so. In this section, we present the theoretical formulation of our bootstrap embedding method on a quantum computer that addresses these challenges. 

Sec. \ref{sec:notation-locality} first set up notations and discuss a few aspects of locality and global symmetry on performing embedding of fermions on quantum computers. Sec. \ref{sec:naive-linear-matching} discuss a naive extension of the classical BE algorithm on quantum computers by matching individual elements of the RDMs directly, and highlight the disadvantage of doing so. Sec. \ref{sec:swap-test-matching} introduces the $\mathtt{SWAP}$ test circuit and show that it achieves the matching between two RDMs coherently. In \ref{sec:quadratic-penalty-opt}, we discuss some subtleties on why it is impossible to incorporate this coherent matching condition into the Lagrange multiplier optimization method, and present an alternative quadratic penalty method to perform the optimization.

\subsection{Fermion-Qubit Mapping - Global Symmetry vs. Locality}
\label{sec:notation-locality}

When mapping electronic structure problem to qubits on quantum computers, it is well-known that the global anti-symmetric property of fermionic wave functions necessarily leads to an overhead in operator lengths or qubit counts \cite{tranter2018comparison}. On the other hand, chemical information is usually local if represented using localized single-particle orbitals \cite{edmiston1963localized,wannier1962dynamics}. In the case of performing bootstrap embedding, this tension between locality of chemical information and global fermionic anti-symmetry is more subtle.
Because bootstrap embedding intrinsically uses the fermionic occupation number in the local orbitals (LOs) to perform matching, it is therefore convenient to preserve such locality when constructing the mapping.
Throughout the discussion, without loss of generality, we assume a mapping that preserves fermionic local occupation number, such as the Jordan-Wigner mapping where each spin-orbital is mapped to one qubit. Our discussion equally applies to cases where a non-local mapping is used (such as parity mapping). In that case, a unitary transformation from the non-local mapping to a local mapping will be required before actually computing the matching conditions.

It is possible to formulate QBE using matching conditions on either qubit reduced density matrices (RDMs) \cite{nielsen2002quantum} or $k$-electron RDMs \cite{mazziotti2012two} for all $k$, both with an exponential number of matrix elements. For simplicity, in the present work we use qubit RDMs in our QBE and leave an efficient formulation in terms of fermionic $k$-electron RDMs for future work. The full density matrix of fragment $A$ is thus provided by $\rho^{(A)} = \ket{\Psi_A}\bra{\Psi_A}$. Given an orbital set $ R \subset O^{(A)}$ for $O^{(A)}$ being set of orbitals in fragment $A$. Let $\rho_R^{(A)}$ signify the RDM obtained from $\rho^{(A)}$ by tracing out the set of qubits 
not in $R$.
Specially, if $R$ only contains orbitals on the edge (center) of fragment $A$, then $\rho_R^{(A)}$ represents information about the density information (for example the occupation number) on the edge (center) of $A$. 

These RDMs can be expanded under an arbitrary set of orthonormal basis $\{ \Sigma_\alpha \}$ as follows
\begin{align}
    \rho^{(A)}_R &= \frac{I + \sum_{\alpha=1}^{4^m-1} \langle \Sigma_\alpha \rangle_A ~ \Sigma_\alpha}{2^m}
    \label{rdm-def}
\end{align}
where $ \langle \Sigma_\alpha \rangle_A = \bra{\Psi_A} \Sigma_\alpha \ket{\Psi_A} = \Tr[\rho^{(A)} ~\Sigma_\alpha], ~\forall\alpha \in [1, 4^m-1] $, and $m = |R|$ is the number of orbitals in the set $R$. One convenient orthonormal basis set is the generalized Gell-Mann basis \cite{bertlmann2008block}. In the special case of a 1-qubit RDM, $\{ \Sigma_\alpha \}$ ($\alpha = x, y, z$) is the familiar Pauli matrices.

\subsection{Naive RDM Linear Matching and its Disadvantage}
\label{sec:naive-linear-matching}

A naive implementation of BE on a quantum computer is to simply replace 1-RDM in Eq. \eqref{1-rdm-def} with the qubit RDM in Eq. \eqref{rdm-def} on the fragment overlapping regions. 
Such an extension imposes matching constraints on each elements of the RDMs, resulting the following constraint vector in analogous to Eq. \eqref{eq:1rdm-constraint}
\begin{align}
    \bm{\mathcal{Q}_{lin}}(\rho^{(A)}_R; \rho^{(B)}_R) = 
    \begin{bmatrix}
        \langle \Sigma_1 \rangle_A - \langle \Sigma_1 \rangle_B \\
        \vdots \\
        \langle \Sigma_{4^m-1} \rangle_A - \langle \Sigma_{4^m-1} \rangle_B
    \end{bmatrix} 
    = \bm{0}.
    \label{eq:linear-constraint}
\end{align}
It is obvious that $\rho_R^{(A)} - \rho_R^{(B)} = 0$, if and only if all the $(4^m-1)$ components in the above constraint are satisfied.

Similarly, we can associate a scalar Lagrange multiplier to each constraint in Eq. \eqref{eq:linear-constraint} and use this linear RDM constraint in place of the 1-RDM constraint $ \bm{\mathcal{Q}_{\textbf{1-RDM}}}(\Psi^{(A)}; \mathbf{P}^{(B)})$ in Eq. \eqref{eq:be_lagrangian_a}. Finding the stationary point of this new Lagrangian gives the same eigenvalue equation as Eq. \eqref{eig-equation} with a new BE potential given by
\begin{align}
    &V_{\rm BE} = \sum_{B\ne A, \mathbb{C}_B \cap \mathbb{E}_A \ne \emptyset}
        \bm{\lambda_B^{(A)}} \cdot \left[ I \otimes \bm{\Sigma_r} \otimes I \right]
    \label{eq:be-potential}
\end{align}
where $\bm{\Sigma_r} = \begin{bmatrix} \Sigma_1, \cdots, \Sigma_\alpha, \cdots, \Sigma_{4^m-1} \end{bmatrix}$ is a $(4^m-1)$-dimensional vector of the orthonormal basis in Eq. \eqref{rdm-def}, and $\bm{\lambda_B^{(A)}}$ is the Lagrange multipliers now modulating the local potentials on each qubit basis, and $n$ is the number of overlapping sites between $A$ and $B$.

To perform the optimization, the eigenvalue equation Eq. \eqref{eig-equation} with the above new BE potential in \eqref{eq:be-potential} can be solved on a quantum computer to obtain an updated wave function for fragment $A$. By iteratively solving the eigenvalue equation and updating the Lagrange multipliers $\{ \bm{\lambda}, \mu\}$ using either gradient-based or gradient-free methods \cite{conn2009introduction}, an algorithm can be formulated to solve the optimization problem. For completeness, we document the algorithm from the naive linear matching of RDMs in Sec. \ref{app:linear-alg} of the SI.

The above is a convenient way to impose the constraint on quantum computers, but it is computationally costly as the number of constraints in \eqref{eq:linear-constraint} increases exponentially as the number of overlapping sites $n$ on neighboring fragments. For each constraint equation, the expectation values $\langle \Sigma_\alpha \rangle$ has to be measured on the quantum computer, which therefore introduces an exponential overhead on the sampling complexity.

In the next section, we introduce a simple alternative to evaluate the mismatch between two RDMs on a quantum computer much faster based on a $\mathtt{SWAP}$ test.

\subsection{Coherent Quantum Matching from \texorpdfstring{$\mathtt{SWAP}$}{swap} Test}
\label{sec:swap-test-matching}

The wave functions of two overlapping fragments are stored coherently as many amplitudes that suppose with each other. The beauty of quantum computers and algorithms lies at the ability to coherently manipulating such amplitudes simultaneously. We may naturally ask: are there quantum algorithms or circuits that can coherently achieve matching between an exponentially large number of amplitudes, without explicitly measuring each amplitude?

\begin{figure}[ht!]
    \centering
    \begin{center}
            \mbox{
                \Qcircuit @C=1em @R=0.7em @!R {
                |0\rangle\quad\quad         & \gate{H} & \ctrl{2} & \gate{H} & \meter & M \\
                |\phi\rangle \quad\quad & \qw  & \qswap & \qw & \qw & \\
                |\psi \rangle \quad\quad      & \qw        & \qswap  & \qw & \qw     & \quad \quad \quad \quad \quad \quad \\
                }
                }
            \end{center}
    \caption{Quantum circuit of a $\swap$ test between two qubits (lower, with state $\ket{\phi}$ and $\ket{\psi}$). The circuit is composed of two Hadamard gate ($H$), a controlled-$\swap$ operation in between, and a final $Z$-basis measurement $M$ on an additional ancilla qubit (top), where $M = 0, 1$.
    }
    \label{fig:swap-test}
\end{figure}
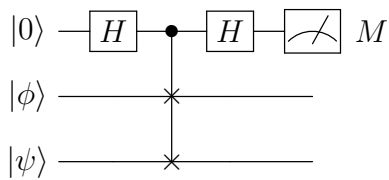

In quantum information, there is a class of quantum protocols to perform the task of estimating the overlap between two wave functions or RDMs under various assumptions \cite{Fanizza2020beyond}. Among these protocols, the $\mathtt{SWAP}$ test is widely used \cite{harrow2013testing,buhrman2001quantum}. Such a $\swap$ test on a quantum computer can also be naturally implemented by simple controlled-$\mathtt{SWAP}$ operations as in Fig. \ref{fig:swap-test}, showing a $\swap$ test between two qubits. The essence of a \texttt{SWAP} test is to entangle the symmetric and anti-symmetric subspaces of the two quantum states ($\ket{\phi}$ and $\ket{\psi}$) to a single ancillary qubit, such that the quantum state of the system before the final measurement is
\begin{align}
    \ket{\Psi} = \frac{1}{2} \left[ \rule{0pt}{2.4ex} \ket{0} \left ( \rule{0pt}{2.4ex} \ket{\phi}\ket{\psi} + \ket{\psi}\ket{\phi} \right) + \ket{1} \left( \rule{0pt}{2.4ex} \ket{\phi}\ket{\psi} - \ket{\psi}\ket{\phi} \right) \right].
\end{align}
By measuring the top \emph{single} ancillary qubit in the usual computational $Z$-basis (collapsing it to either the $\ket{0}$ or $\ket{1}$ state), the overlap of the two qubit wave function, $|\langle \phi| \psi \rangle|$, can be directly obtained from the measurement outcome probability:
\begin{align}
    {\rm Prob}[M = 0] = \frac{1 + |\langle{\phi}| \psi \rangle|^2}{2},
\end{align}
without requiring explicit estimation of the density matrix elements of each individual qubit. 

Can we recast the linear matching conditions as linear combination of several $\mathtt{SWAP}$ tests?
Observe that an equivalent condition alternative to Eq.~\eqref{eq:linear-constraint} is the following quadratic matching condition
\begin{align}
    \mathcal{Q}_{quad}(\rho^{(A)}_R; \rho^{(B)}_R) = \Tr[\left( \rho^{(A)}_R - \rho^{(B)}_R \right)^2] = 0.
    \label{eq:quadratic-constraint}
\end{align}
Interestingly, the above quadratic constraint can be rewritten as a linear combination of three different multi-qubit generalization of the $\mathtt{SWAP}$ tests (with each repeated multiple times), regardless of the number of overlapping sites (Fig. \ref{fig:be-schematic}iiiq).
Two of the $\mathtt{SWAP}$ tests are to estimate the purity of $\rho^{(A)}_R$ and $\rho^{(B)}_R$ each, while the third one is to estimate the overlap between $\rho^{(A)}_R$ and $\rho^{(B)}_R$. See SI Sec. \ref{app:constraint-equivalent-proof} for a proof of the equivalence between the two quantum matching conditions) and Sec. \ref{app:swap-test} on how to generalize the $\swap$ test on two qubits to a multi-qubit setting and how to relate the $\mathtt{SWAP}$ test results to the quadratic constraint.

The reformulation of the quadratic constraint allows us to estimate the mismatch between two fragments by measuring only a single ancilla qubit (estimating three different amplitudes). As compared to the linear constraint case where an exponentially large number of constraints have to be estimated individually ($4^m-1$ where $m = |R|$ is the number of overlapping sites again), the quadratic matching based on $\mathtt{SWAP}$ tests achieves an exponential saving in the types of measurements required. 

Furthermore, the reduction of the mismatch to the estimation of only a few (three) amplitudes in $\mathtt{SWAP}$ tests allows an additional quadratic speedup by amplifying the amplitude of the ancilla qubit before measure it. We will discuss more details on how to achieve the quadratic speedup in Sec. \ref{sec:aa-quadratic-speedup}. Admittedly, such amplitude amplification algorithm may be applied even to the naive linear RDM matching by boosting individual RDM amplitude, but the resulting quantum circuit will be much more complicated.

\subsection{Optimization Using the Quadratic Penalty Method}
\label{sec:quadratic-penalty-opt}

With an efficient way to estimate the quadratic penalty constraint established in Eq. \eqref{eq:quadratic-constraint}, it now appears feasible to use this new constraint in Eq. \eqref{eq:be_lagrangian_a} as in the case of linear constraint. However, the nature of the quadratic matching in Eq. \eqref{eq:quadratic-constraint} makes the same Lagrange multiplier optimization method used in the linear case invalid. We first discuss in more detail why this approach fails, in Sec. \ref{sec:violation_constraint_qualif}; we then describe an alternative way of treating the quadratic constraint as a penalty term to optimize the resulting objective function in Sec. \ref{sec:detail-quadratic-penalty}.

\subsubsection{Violation of the Constraint Qualification}
\label{sec:violation_constraint_qualif}

A necessary condition to use the Lagrange multiplier method for constraint optimization is that the gradient of the constraint itself with respect to system variables has to be non-zero at the solution point (this guarantees a non-zero effective potential to be added to the original Hamiltonian), a.k.a., constraint qualification \cite{mangasarian1967fritz,bertsekas2016nonlinear}. Specifically, we require $\nabla \mathcal{Q}_{quad}(\rho^{(A)}_R; \rho^{(B)}_R) \ne 0$ when $\rho^{(A)}_R = \rho^{(B)}_R$. 

Unfortunately, in the quadratic case, we have 
\begin{align}
    \nabla \mathcal{Q}_{quad}(\rho^{(A)}_R; \rho^{(B)}_R) \propto \rho^{(A)}_R - \rho^{(B)}_R = 0
\end{align} 
when $\rho^{(A)}_R$ and $\rho^{(B)}_R$ matches, which violates the above condition. Note that any high-order constraint other than linear order will violate the constraint qualification. The existence of such constraint qualification makes sense also from a physical point of view. Because the gradient $\nabla \mathcal{Q}_{quad}(\rho^{(A)}_R; \rho^{(B)}_R)$ enters the eigenvalue equation \eqref{eq:VBE} as the BE potential $V_{\rm BE}$ modulated by the Lagrange multipliers. The vanishing of this potential near the solution point means there is no way to modulate $V_{\rm BE}$ by adjusting the Lagrange multipliers, and therefore will lead to failure of convergence of the Lagrange multiplier.

Alternatively, the quadratic constraint can be treated as a penalty by using $\lambda_B^{(A)} \mathcal{Q}_{quad}(\rho^{(A)}_R; \rho^{(B)}_R)$ to substitute the constraint $\bm{\lambda^{(A)}_{B}} \cdot \bm{\mathcal{Q}_{\textbf{1-RDM}}}(\Psi^{(A)}; \mathbf{P}^{(B)})$ in Eq. \eqref{eq:be_lagrangian_a}. We can then employ the quadratic penalty method \cite{staszczak2010augmented} to minimize this cost function. To highlight the distinction of quadratic penalty method from the Lagrange multiplier method, we use ``cost function" instead of ``Lagrangian" to refer to the objective function in the quadratic penalty case.

\subsubsection{Details of the Quadratic Penalty Method}
\label{sec:detail-quadratic-penalty}

The idea of the penalty method is to use the constraint as a penalty where the magnitude of $\lambda_B^{(A)}$ serves as a weight to the penalty. Initially, $\lambda_B^{(A)}$ is set to a small constant, and then we treat the resulting cost function as an unconstrained minimization where its minimum is found by varying the wave functions. 
The next step is to increase $\lambda_B^{(A)}$ to a larger value leading to a new Lagrangian, which is then minimize again by varying the wave function parameters. This procedure is repeated until the penalty parameter $\lambda_B^{(A)}$ is large enough to guarantee a small mismatch $\mathcal{Q}_{quad}(\rho^{(A)}_r; \rho^{(B)}_r)$. In our case, we choose all $\lambda_B^{(A)} = \lambda$ for all pairs of adjacent fragments.

It is helpful to note that optimization of the wave function is done again using the eigenvalue equation as in Eq. \eqref{eig-equation} by tuning the BE potential $V_{\rm BE}$. In other words, for a fixed penalty parameter $\lambda$, the fragment Lagrangian $\mathcal{L}_A(\{V_{\rm BE}\})$ is minimized with respect to $V_{\rm BE}$. 
For a particular parametrization in terms of local potentials $\{ v_\alpha \}$ on the edge sites of fragment A
\begin{align}
    V_{\rm BE}(\{v_\alpha\}) = \sum_{\alpha = 0}^{M} v_\alpha ~ I \otimes \Sigma_\alpha \otimes I,
    \label{vbe-expression}
\end{align}
where $\{\Sigma_\alpha\}$ is a set of Hermitian generator basis of size $M$ on the edge sites of fragment A (can be Pauli operators for a single edge site), and $\{v_\alpha\}$ is the corresponding local potential (real numbers). Note that $M$ in Eq. \eqref{vbe-expression} can be much smaller than the total number of generators ($4^m$) on the edge sites, because in each bootstrap embedding iteration, only a small local potential is added to the Hamiltonian. This perturbative nature of the bootstrap embedding iteration allows us to expand the BE potential $V_{\rm BE}$ in each iteration under the Hermitian generator basis from the previous iteration, such that the BE potential in each iteration is diagonal dominant, i.e., $M \ll 4^m$ where $n$ is the number of edge sites on any fragment $A$.

To update $\{v_\alpha\}$, we derive the following gradient (SI Sec. \ref{app:gradient-quadratic})
\begin{align}
    \frac{d\mathcal{L}^{(A)}}{dv_\alpha}
    &= \sum_{n' \ne 0} \left[
    \mathbf{C}^\dagger (\mathbf{I \otimes W_\alpha^{(n')} \otimes I}) \mathbf{C^{(n')}} \right] \times \left[ \mathbf{C^{(n')\dagger}} \left( \mathbf{H^{(A)}} +  \mathbf{\mathcal{E}_0^{(A)}}  + 2 \lambda  \left( \mathbf{I \otimes ( \rho_{\mathbb{E}_A} - \rho_{\mathbb{C}_B}) \otimes I} \right) \right) \mathbf{C} \right]
    \label{quadratic-gradient}
\end{align}
$\forall \alpha \in [0,M]$,
that can, in principle, be used to perform the updating of $V_{\rm BE}$ to minimize $\mathcal{L}^{(A)}$. In the above, $\mathbf{C^{(n)}}$ is the eigenvector of the $n$-th eigenstate ($n \ge 1$) while $\mathbf{C}$ is the eigenvector of the ground state, $\mathbf{W_\alpha^{(n')}}$ is a perturbation matrix between ground state and the $n'$-th eigenstate for the $\alpha$-th Pauli basis at the edge site of fragment $A$, whereas $\rho_{\mathbb{E}_A}$ and $\rho_{\mathbb{C}_B}$ are the RDM at the edge and center sites of fragment $A$ and $B$, respectively. 

The above gradient in Eq. \eqref{quadratic-gradient} is only formally useful, but computing it exactly requires all the eigenstates to be known (not only the ground state) which is clearly very costly if possible. Nevertheless, it serves as a good starting point to develop \emph{approximated} updating scheme or to perform bootstrap embedding for excited states. We leave such topics for future investigation. In the present work, instead of using Eq. \eqref{quadratic-gradient} to update $V_{\rm BE}$, we employ gradient-free schemes to update $\{ v_\alpha \}$ and measure the required expectation values using $\mathtt{SWAP}$ test to obtain the mismatch to evaluate the cost function $\mathcal{L}^{(A)}$.

We note that one additional advantage of this quadratic penalty method is that it can be easily integrated with variational eigensolvers \cite{tilly2022variational} by treating the quadratic penalty as an additional term in the VQE cost function \cite{kuroiwa2021penalty}. The drawback is that the optimized wave function only \emph{exactly} equals to the true wave function when the penalty goes to infinity $\lambda \rightarrow \infty$. Practically, we find that choosing the penalty parameter large enough is sufficient to obtain satisfactory results.

\section{Quantum Bootstrap Embedding Algorithms}
\label{sec:qbe-algorithms}

Given the theoretical formulation of QBE method in Sec. \ref{sec:qbe-methods}, we present a general hybrid quantum-classical algorithm in this section that can be practically used to solve the BE problem on quantum computers to find the BE potentials $V_{\rm BE}$ that satisfies the matching condition. 

In our quantum bootstrap embedding algorithm, the electronic structure problem of the total system is formulated as a minimization of a composite objective function with a penalty term constructed from the matching conditions on the full qubit RDMs on overlapping regions of adjacent fragments. We then design an iterative hybrid quantum-classical algorithm to solve the optimization problem, where a quantum subroutine as an eigensolver is employed to prepare the ground state of fragment Hamiltonian. The quantum matching algorithm employs a $\mathtt{SWAP}$ test \cite{barenco1997stabilization,buhrman2001quantum} between wave functions of two fragments to evaluate the matching conditions, which is a dramatic improvement as compared to the straightforward method of measuring an exponential number (with respect to the number of qubits on the fragment edge) of RDM elements. Additionally,
the quantum bootstrap embedding framework is internally self-consistent without the need to match fragment density matrices to external more accurate solutions.
The adaptive sampling changes the number of samples as the optimization proceeds in order to achieve an increasingly better matching conditions. We note that the $\swap$ test adds only little computational cost to quantum eigensolvers which can be readily performed on current NISQ devices. The amplitude amplified coherent quantum matching requires iterative application of eigensolvers multiple times which are more suitable for small fault-tolerant quantum computers.

The rest of this section is organized as follows. Sec. \ref{sec:main-sub-algorithm} gives an outline of the QBE algorithm with the quadratic penalty method. Sec. \ref{sec:alg-eigensolver} discusses possible choices of quantum eigensolvers with an analysis on sampling complexities. We then present a way to achieve an additional quadratic speedup by using coherent amplitude estimating algorithm in Sec. \ref{sec:aa-quadratic-speedup}.

\subsection{The Algorithm}
\label{sec:main-sub-algorithm}

We present a high-level framework of the main algorithm in this section. As a comparison, the QBE algorithm with naive linear matching can be found in SI Sec. \ref{app:linear-alg}. Code for the algorithms and data for generating the plots are available as open source on github \cite{QBE2022}.

To quantify the mismatch across all fragments, we define $\Delta\rho$ to be the root mean square density matrix mismatch averaged over all the overlapping sites of all the fragments according to 
\begin{equation}
    \Delta\rho = \sqrt{\frac{1}{N_{sites}}\sum_{A,B}\sum_{r \in \mathbb{E}^{(A)} \cap \mathbb{C}^{(B)}} \Tr[\left(\rho^{(B)}_r - \rho^{(A)}_r \right)^2]}
    \label{mismath-delta-rho}
\end{equation}
where $\Tr[\left(\rho^{(B)}_r - \rho^{(A)}_r \right)^2] = \mathcal{Q}_{quad}(\rho^{(A)}_r; \rho^{(B)}_r)$ as in Eq. \eqref{eq:quadratic-constraint}, which may also be recognized as the Frobenius norm of $(\rho^{(B)}_r - \rho^{(A)}_r )$. $N_{sites}$ is the total number of terms in the double sum in Eq. \eqref{mismath-delta-rho}, $N_{sites} = \sum_{A \ne B} | \mathbb{E}^{(A)}\cap \mathbb{C}^{(B)} |$,
with $|\mathcal{S}|$ denoting the number of elements in set $\mathcal{S}$.

The cost function $ \mathcal{L}^{(A)}(\lambda)$ being optimized is discussed in Sec. \ref{sec:violation_constraint_qualif}. For clarity, we write it explicitly here
\begin{align}
    \mathcal{L}^{(A)}(\lambda) =& \langle \hat{H}^{(A)} \rangle_A 
    + \sum_{B} \lambda \mathcal{Q}_{quad}(\rho^{(A)}_R; \rho^{(B)}_R)
    \label{eq:cost_function_alg},
\end{align}
with $\mathcal{Q}_{quad}$ given by Eq. \eqref{eq:quadratic-constraint}. We have omitted the term $\mathcal{E}^{(A)}$ for simplicity since the normalization of the wave function is guaranteed for a fault-tolerant quantum computer. However, this term can be important on a noisy quantum computer where the purity of the wave function can be contaminated. Note the expectation value in Eq. \eqref{eq:cost_function_alg} has to be estimated by collecting samples on a quantum computer.

The quantum bootstrap embedding algorithm with quadratic penalty method is presented below in Alg. \ref{alg:qbe_quad}. The algorithm takes as its input the total Hamiltonian of the original system, and then perform the fragmentation and parameter initialization, followed by the main optimization loop to achieve the matching. Finally, it returns the optimized BE potential $V_{\rm BE}^{(A)}$ for any fragment $A$ and the final mismatch $\Delta\rho$. Inside the main loop (line 9 of Alg. \ref{alg:qbe_quad}), the cost function $ \mathcal{L}^{(A)}(\lambda)$ for each fragment $A$ is minimized for a fixed penalty parameter $\lambda$ (line 10 and 11). The penalty $\lambda$ is then increased geometrically (line 12) until the mismatch criteria is met, i.e., $\Delta \rho \le \varepsilon$.

\begin{algorithm}[H]
    \caption{
   Quantum bootstrap embedding algorithm: quadratic penalty method
    }\label{alg:qbe_quad}
    \textbf{Input}: Geometry of the total molecular system and the associated \textit{ab initio} Hamiltonian. \\
    
    \texttt{\\}
    \tcc{Initialization}
    \textbf{Fragmentation}: Divide the full molecular system into $N_{frag}$ overlapping fragments;

    \For{A = 1 to $N_{frag}$}{
        Generate $H^{(A)}$ using Eq. \eqref{embedding-projection-int} of SI Sec. \ref{app:frag};
        
        Set $V_{\rm BE}^{(A)} = 0$;
    }
    \textbf{Parameter initialization:} 
    set initial penalty factor $\lambda = 1$; set initial mismatch $\Delta\rho > \epsilon$.
     
    \texttt{\\}
    \tcc{Main loop:}
    \While{$\Delta\rho > \varepsilon$}
    {
        \For{A = 1 to $N_{frag}$}
        {
           \textbf{Minimize $ \mathcal{L}^{(A)}(\lambda)$ as in Eq. \eqref{eq:cost_function_alg} :}
           Repeatedly generate $V_{\rm BE}^{(A)}$ and estimate the penalty loss function $ \mathcal{L}^{(A)}(\lambda)$ using $\swap$ test.
        } 
        \textbf{Increase penalty parameter:} 
        $\lambda \leftarrow \gamma\lambda$, for some fixed $\gamma > 1$.\\

        \textbf{Update mismatch}:
        \For{A = 1, $N_{frag}$}
        {
            Estimate $\mathcal{Q}_{quad}(\rho^{(A)}_r; \rho^{(B)}_r)$ using $N_{samp}^{\swap}$ (Eq. \eqref{eq:nsamp_swap}) samples for each $\swap$ test.
        }
        Classically compute the mismatch $\Delta\rho$ using Eq. \eqref{mismath-delta-rho}.
    }
    \textbf{Returns:}
    $\left(H^{(A)} + V_{\rm BE}^{(A)}\right)$ for all $A$, $\Delta\rho$. 
\end{algorithm}

A key step of the algorithm is the minimization of $\mathcal{L}^{(A)}(\lambda)$ at line 11, which consists of repeatedly generating the BE potential $V_{\rm BE}^{(A)}$ and estimate the mismatch using $\swap$ test. BE potentials $V_{\rm BE}^{(A)}$ are generated differently for different optimization algorithms. In our implementation, a quasi-Newton method, the L-BFGS-B \cite{byrd1995limited} algorithm, is used at line 11 for minimizing $\mathcal{L}^{(A)}(\lambda)$, where $V_{\rm BE}^{(A)}$ is proposed by the optimizer in order to estimate the inverse Hessian matrix to steer the optimization properly. Alternatively, if derivative-free methods such as Nelder-Mead \cite{nelder1965simplex} is used, $V_{\rm BE}^{(A)}$ will be generated in a high-dimensional simplex defined by the coefficients $\{ v_\alpha \}$ in Eq. \eqref{vbe-expression}, which is repeatedly refined.

Once $V_{\rm BE}^{(A)}$ is generated, the first term in the cost function in Eq. \eqref{eq:cost_function_alg} is estimated by invoking the quantum eigensolver for the Hamiltonian $\left(H^{(A)} + V_{\rm BE}^{(A)}\right)$. The second term, the mismatch in Eq. \eqref{eq:cost_function_alg} can be estimated by measurement outcomes of the ancilla qubit in the $\swap$ test (Sec. \ref{sec:qbe-methods}). The mismatch estimation at line 13 is performed in the same way as those in line 11. Note that the number of samples $N_{samp}^{\swap}$ (Eq. \eqref{eq:nsamp_swap}) for the $\swap$ test estimation can be changed adaptively in different BE iterations for different accuracy, which we discuss in detail in the next section.

\subsection{Eigensolver Subroutines and Sampling Complexity}
\label{sec:alg-eigensolver}

Two major quantum eigensolvers, QPE \cite{svore2013faster} and VQE \cite{tilly2022variational} can be used in line 11 and 14 of Alg. \ref{alg:qbe_quad} to estimate the cost function. QPE is an exact eigensolver, where the system wave function collapses to the exact ground state regardless of the number of evaluation qubits used. In contrast to QPE, VQE is an approximate eigensolver and the results depends on the choice of ansatz and the optimization algorithm used.

A crucial feature of a quantum eigensolver is its probabilistic nature, in a sense that any measurement collapses the entire quantum state. This perspective allows us to treat a quantum eigensolver as a sign-problem-free sampling oracle for correlated electronic structure problems where Ref. \cite{huggins2022unbiasing} provides a concrete example.

The stochastic nature also means a more careful treatment on the number of samples is required to fully quantify any potential quantum speedup. In general, for typical iterative mixed quantum-classical algorithms, some parameters are usually passed from one iteration to the next, where the parameters are estimated by repeatedly sampling from a quantum eigensolver oracle through proper measurement. This means the uncertainty on these parameters estimated from one iteration has to be small enough to avoid a divergence of the algorithm as iteration continues.

In particular in the bootstrap embedding case, the sampling accuracy on the fragment overlap of each iteration has to be good enough such that the uncertainty of the mismatch passed to the next iteration will not spoil the iteration and lead to diverging results as iterations continue. In the following, sampling complexities of classical matching and $\swap$-test-based quantum matching are compared.

When estimating the overlap $S$ to an accuracy $\epsilon$ naively by density matrix tomography (TMG) of individual RDM elements, it is shown under mild assumptions that the total number of samples required (Sec. \ref{app:sample-advantage-lin-quad} of SI)
\begin{align}
    N_{samp}^{\rm TMG}(S, \epsilon, n) = \mathcal{O} (e^n) \left( \frac{D}{\epsilon^2} \right),
    \label{nsamp_tomography}
\end{align}
where $n$ is the number of qubits on the overlapping region, and $D$ is a system-dependent constant as a function of the two RDMs.
In contrast, the quantum matching based on $\swap$ test costs
\begin{align}
    N_{samp}^{\swap}(S, \epsilon) = \left( \frac{1-S^2}{8} \right) \frac{1}{\epsilon^2},
    \label{eq:nsamp_swap}
\end{align}
which is independent of the size $n$ of the overlapping region of two fragments. This demonstrates that our quadratic quantum matching achieves an exponential speedup compared to naive tomography of density matrices. This dramatic speedup is perhaps not that surprising because we only care about one particular observable (the overlap) instead of the full subsystem RDMs. Therefore, if the observable can be mapped to measurement outcome of few qubits by some quantum operations ($\swap$ test in this case), advantages are expected in general.

Moreover, the dependence of $N_{samp}^{\swap}(S, \epsilon)$ on the overlap $S$ and estimation accuracy $\epsilon$ allows an adaptive sampling schedule to be implemented for line 11 and 14 of Alg. \ref{alg:qbe_quad}. For example, we may use the overlap $S$ estimated from the previous BE iteration to compute the required $N_{samp}^{\swap}$ in the current BE iteration. The accuracy $\epsilon$ can also be dynamically tuned according to the error of the first term in Eq. \eqref{eq:cost_function_alg}, as well as the value of the penalty parameter $\lambda$. For example, at the beginning BE iterations, the mismatch ($\Delta\rho$ or more precisely $\mathcal{Q}_{quad}(\rho^{(A)}_r; \rho^{(B)}_r)$) is large so that a moderate $\epsilon$ suffices. As the BE iteration proceeds, the overlap converges exponentially, therefore an exponentially decreasing $\epsilon$ has to be used as well. A numerical value of $\epsilon$ needs be determined from case to case.

In addition, Eq. \eqref{eq:nsamp_swap} suggests an interesting behavior. As the QBE algorithm proceeds and the overlap $S$ increases, fewer samples are needed to achieve a target accuracy. If $S$ approaches 1 exponentially fast as $S \sim 1 - e^{- \gamma \cdot n_{\text{iter}}}$ for some constant $\gamma$, then the required number of samples for $\swap$ will decrease exponentially as BE iteration $n_{\rm iter}$ goes $N_{samp}^{\swap} \sim e^{- \gamma \cdot n_{\text{iter}}} / \epsilon^2$. In practice, the overlap of two subsystem can never approach 1 but saturates to a constant $0<c<1$ when matching is achieved, and therefore $N_{samp}^{\swap} \sim (1-c)/\epsilon^2$ still obeys the $1/\epsilon^2$ scaling generally. This, on the other hand, suggests that a larger overlapping region is advantageous to reduce $N_{samp}^{\swap}$ because the RDM of a larger subsystem of a pure state will have greater purity (hence larger $c$) in general.

\subsection{Additional Quadratic Speedup}
\label{sec:aa-quadratic-speedup}

The core of many quantum speedups over classical algorithms lie at the ability of quantum computers to directly manipulate the \emph{probability amplitude} instead of probability itself, while classical computers only have access to probability. With this idea, the above perspective of treating a quantum eigensolver as an oracle where some amplitude is estimated through proper measurements allows us to achieve an additional quadratic speedup in our quantum bootstrap embedding algorithm. This section compares two different versions of quantum matching algorithms in QBE, the $\swap$ and the $\swap+$AE algorithms. However, the same argument of quadratic speedup applies to classical sampling based eigensolvers such as VMC as discussed in detail at the end of this section.

The intuition is that instead of directly measuring a small quantum amplitude to accumulate enough counts to reduce the error bar, we may use quantum algorithms to first amplify the amplitude before the measurement. One way of understanding this is that Eq. \eqref{eq:nsamp_swap} contains an overlap-dependent prefactor $(1-S^2)$ as discussed above. If the overlap $S$ (as a probability amplitude) can be manipulated on the quantum computer \emph{easily} such that $(1-S^2)$ is on the order of $\epsilon$, then $N_{samp}^{\swap}$ will be proportional to only $1/\epsilon$ instead of $1/\epsilon^2$. There are well-established ways of performing such amplitude amplification task via coherent quantum algorithms \cite{brassard2002quantum}. See SI Sec. \ref{app:aa+binary-search} for the construction of the amplitude amplification and binary search quantum algorithm.

In particular, in each iteration of the algorithm, it can be shown that by combining oblivious amplitude amplification and a binary search protocol, estimating the overlap up to precision $\epsilon$ between adjacent fragments takes $N_{samp}^{\swap+AE}$ samples (state preparation and $\mathtt{SWAP}$ tests)
\begin{align}
    N_{samp}^{\swap+\rm AE} = \frac{\sqrt{2}}{2\ln(2)\epsilon} \ln^2(\frac{1}{\epsilon}),
    \label{eq:nsamp_swap+bs}
\end{align}
regardless of the overlap $S$.

Comparing \eqref{eq:nsamp_swap+bs} with \eqref{eq:nsamp_swap}, the above analysis suggests that our coherent quantum matching algorithm achieves a quadratic speed up (up to a factor of ${\rm polylog}(\frac{1}{\epsilon})$) as compared to the $\swap$ test based quantum matching algorithm, which is consistent with typical behavior of a Grover-type of search algorithm. Moreover, in contrast to \eqref{nsamp_tomography}, an exponential advantage is present with respect to the size of the overlapping region, indicating the benefit of using our quadratic QBE algorithm for fragment matching in the presence of large overlapping region.

In the above, we leverage amplitude amplification to achieve a quadratic speedup of a quantum subroutine based on $\swap$ test. More generally, such amplitude amplification technique can be utilized to achieve a general quadratic speedup in the required number of samples for any Monte Carlo classical algorithms \cite{wocjan2008speedup,yung2012quantum,montanaro2015quantum}. This can be understood by realizing that classical probability distributions may be encoded in the amplitudes of the quantum state of a quantum computer, where measurements performed after some unitary quantum computation is similar to \emph{sample} from the quantum computer to extract the probability distributions. When treating the unitary quantum computation part as a quantum blackbox, it is then easier to understand the quadratic speedup in the number of samples as compared to classical Monte Carlo methods. In our case, the quantum blackbox is the quantum eigensolver used to find the ground state for each fragment, while the classical blackbox is the stochastic classical eigensolvers such as VMC. 


\section{Results and Discussions}
\label{sec:results}

With the theoretical foundation and algorithms discussed in previous sections, we present numerical results in this section using a typical benchmark system in quantum chemistry, hydrogen chains under minimal basis. In Sec. \ref{sec:results-convergence}, we demonstrate the convergence of the QBE algorithm with an exact solver (at infinite sampling limit) using an H$_8$ molecule with STO-3G basis. In Sec. \ref{sec:results-sample}, we present numerical evidence for the sampling advantages of the QBE algorithm in terms of overlapping fragment size (non-interacting H$_4$ molecule with STO-3G basis) and target precision over incoherent estimation and classical VMC sampling (H$_8$ molecule under STO-3G basis). Numerical results using \emph{approximate} variational quantum eigensolvers (VQE) on a random spin model and a perturbed H$_4$ molecule are documented in Sec. \ref{app:vqe} of SI for interest readers, where a similar BE convergence is established at the beginning iterations but later plateaus, likely due to intrinsic VQE ansatz truncation errors. A detailed discussion of BE+VQE goes beyond the scope of this work which we leave for future investigation.

\subsection{Convergence of QBE in Infinite Sampling Limit}
\label{sec:results-convergence}

\begin{figure}[ht!]
    \centering
    \includegraphics[width=9cm]{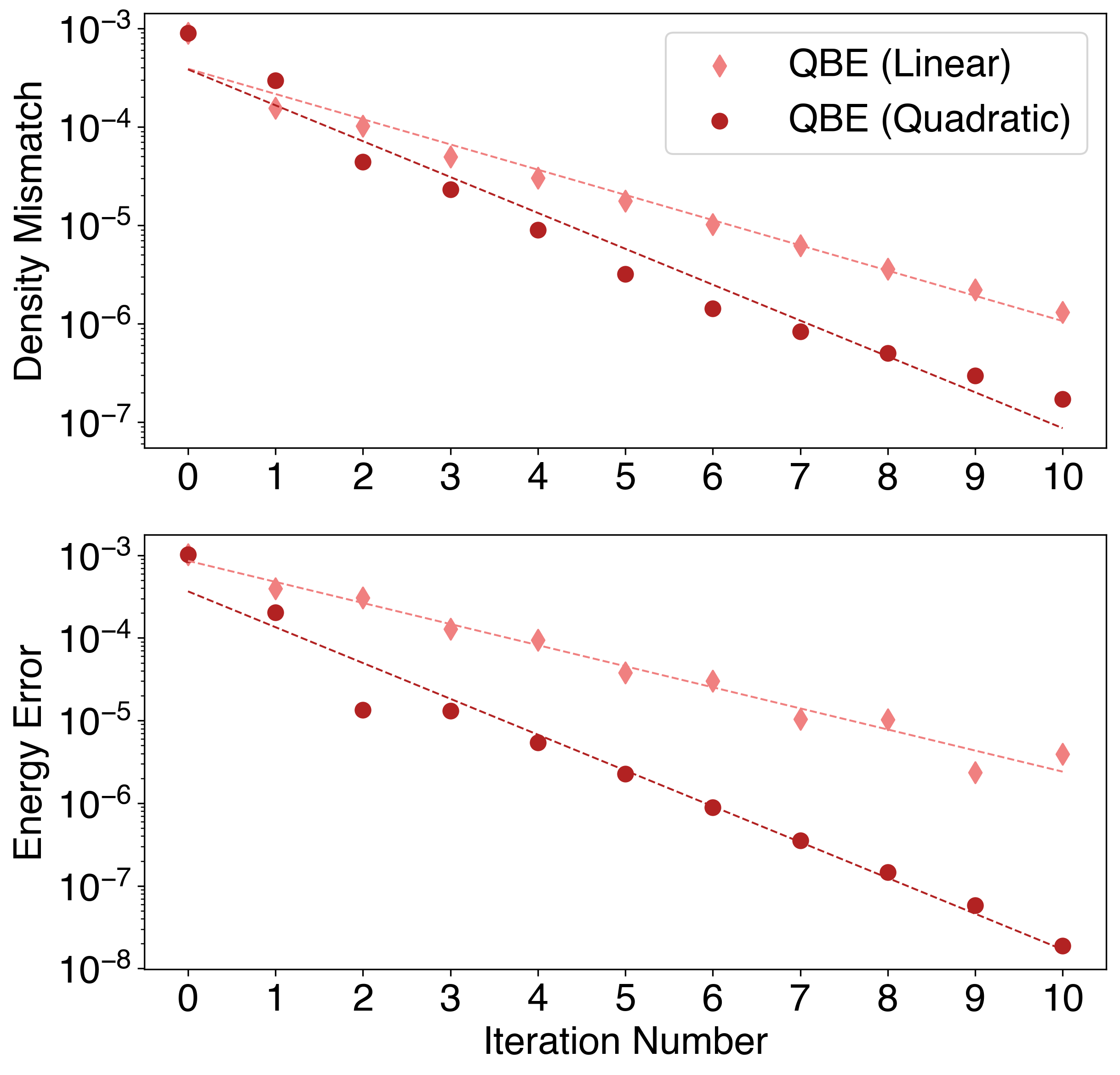}
    \caption{\label{fig:be-convergence-deterministic} Convergence of the quantum bootstrap embedding algorithms on (a) density mismatch and (b) energy error for the linear constraint (pink) and quadratic penalty method (red) in the infinite sample limit for an H$_8$ molecule. The dashed trend lines in both panels indicate an exponential fit. 
    }
\end{figure}

We focus on demonstrating the convergence of QBE in the infinite sampling limit by using exact deterministic solver with the quadratic constraint in Eq. \eqref{eq:quadratic-constraint} and linear constraint in Eq. \eqref{eq:linear-constraint}. 
As a standard benchmark system for electronic structure, we perform QBE on a H$_8$ chain under a minimal STO-3G basis, which is fragmented into six overlapping fragments each with six embedding orbitals. 
Fig. \ref{fig:be-convergence-deterministic}a shows the exponential convergence of the density mismatch for an H$_8$ molecule in both linear and quadratic constraint cases. This convergence behavior of QBE matches the convergence of classical BE in Fig. \ref{fig:classical_baseline} with exact classical solver (FCI), demonstrating the correctness of the new constraints. The agreement on the convergence with classical BE in Fig. \ref{fig:classical_baseline} is expected since at infinite sampling limit, the outer iterations in both classical and quantum BE are the same classical optimization routine.

To quantify how much energy error the final converged result has, Fig. \ref{fig:be-convergence-deterministic}b shows the absolute value of the error in energy using the energy in the last (11$^{th}$) iteration as a reference. 
We can see that the energy errors from both the linear and quadratic constraint algorithm exhibit similar exponential convergence as the density mismatch. Moreover, the energy in both cases converge to the same value within 10$^{-6}$ in the last iteration (not shown in the figure). We note that the linear constraint case shows a slightly oscillatory convergence, while the quadratic case is free of such oscillatory behavior. The fact that quadratic appears to converge slightly faster than linear may be coincidence for the system investigated, and the convergence rate in general depends on the optimization algorithm chosen. See Sec. \ref{app:qbe-calculation} of the SI for a detailed description on definition of the energy.

\subsection{Sampling Advantage of Coherent Quantum Matching}
\label{sec:results-sample}

In the previous section, we have seen that our quantum bootstrap embedding algorithm convergence as expected 
in the infinite sampling limit. It is also seen (in the SI) that the approximate VQE leads to biased behavior on the density matching. 
In practice, only a finite number of samples can be collected on a quantum computer, and we will focus on theis scenario in this section. In particular, we present numerical data demonstrating the sampling advantage of our coherent quantum matching algorithm. Sec. \ref{sec:results-sample-fragment-size} discusses the sampling advantage of the quantum matching algorithm for an overlapping region of increasing size, echoing the analytical sampling complexity derived in Sec. \ref{sec:alg-eigensolver}. In Sec. \ref{sec:results-sample-quadratic-speedup}, the additional quadratic speedup in estimating the overlap via amplitude amplification and binary search (AE) is presented, which agrees with the theoretical sampling complexity in Sec. \ref{sec:aa-quadratic-speedup}.

\subsubsection{Advantage in Fragment Overlap Size}
\label{sec:results-sample-fragment-size}

To perform bootstrap embedding, it is usually advantageous to partition the system into fragments with large overlapping region to increase the convergence rate, because a large overlapping region necessarily means more information is provided to update the local potential for the following BE iteration. However, as is seen in Eq. \eqref{nsamp_tomography} of Sec. \ref{sec:alg-eigensolver}, a larger overlapping size also lead to a potentially exponentially higher sampling complexity versus the number of qubits in the overlapping region if estimating the overlap naively from density matrix tomography (TMG). The quantum matching algorithm implemented by a $\swap$ test (Fig. \ref{fig:be-schematic}iiiq) bypass the need for density matrix tomography, and therefore leads to a sample complexity as in Eq. \eqref{eq:nsamp_swap} independent of the size of the overlapping region.

\begin{figure}[ht!]
    \centering
    \includegraphics[width=9cm]{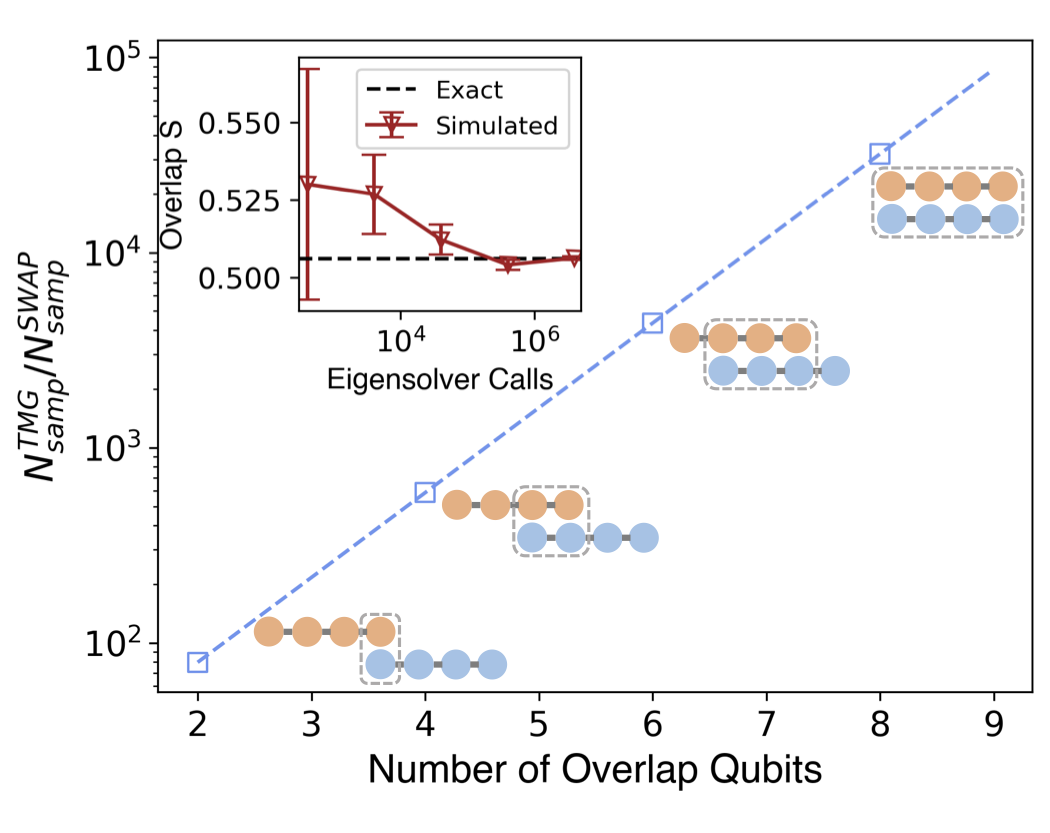}
    \caption{\label{fig:swap-size} Sampling complexity ratio of naive density matrix tomography (TMG) and $\swap$ test versus number of qubits in the overlapping region for a target precision $\epsilon = 0.001$ on overlap $S$. The inset shows a simulated convergence of overlap ($S$) estimation using quantum matching ($\swap$) for the case of two overlapping qubits. Data are obtained from a non-interacting chain of H$_4$ (see SI Sec. \ref{app:swap-test-circuit} for details). 
    }
\end{figure}

To validate our theoretical sample complexity, a simulation of the quantum matching algorithm with QPE as an eigensolver for two identical H$_4$ chain is performed using a noiseless Qiskit AerSimulator (see SI Sec. \ref{app:swap-test-circuit} for more details) for an increasing overlap region ranging from 2 to 4, 6, and 8 qubits (schematic in Fig. \ref{fig:swap-size}). In the simulation, we first use QPE to prepare the ground state for two non-interacting H$_4$ molecules separately. A $\swap$ test is then performed on relevant qubits in the overlapping region between the two H$_4$ molecules. The evaluation qubits for QPE and the ancilla qubit for $\swap$ test are all measured afterwards. Post-selection on the QPE evaluation qubits are performed in order to select the ground states of H$_4$ molecules. The $\swap$ test results are processed and converted to the estimation on the overlap $S$. 

The inset of Fig. \ref{fig:swap-size} shows the estimated overlap $S$ as a function of sample size (number of eigensolver calls) in the case of two overlap qubits. The estimated overlap converges to the exact value (black dashed horizontal line) for roughly four million samples within $5\times10^{-4}$ (error bar invisible for the last data point). This demonstrates the correctness of our quantum matching algorithm.

By repeating similar estimation as described above for increasingly larger overlapping regions, the exponential sampling advantage of the quantum matching algorithm over naive density matrix tomography is evident in Fig. \ref{fig:swap-size}. As we can see, to achieve a constant target precision of $\epsilon = 0.001$ on the overlap $S$, the ratio between the $\swap$ test estimation and the naive tomography estimation for the required number of eigensolver calls increases exponentially as the number of qubits.

We note that in general, overlaps between density matrices are not low-rank observables, so the sampling complexity of estimating it is likely to be high. However, more efficient sampling schemes may exist than the naive density matrix tomography as presented in Eq. \eqref{nsamp_tomography}. For example, by sampling the differences in the RDMs between the current and the previous BE iterations, the sampling complexity could be much better than exponential. We leave this for future investigation.

\subsubsection{Additional Quadratic Speedup in Accuracy} 
\label{sec:results-sample-quadratic-speedup}

We have seen in the previous section that the quantum matching implemented by a $\swap$ test shows a potentially exponential sampling advantage in terms of the size of the overlapping region as compared to naive density matrix tomography (compare Eqs. \eqref{eq:nsamp_swap} to \eqref{nsamp_tomography}). However, the sample complexity in the estimation accuracy $\epsilon$ follows the same scaling of $1/\epsilon^2$ as classical sampling based eigensolvers such as VMC. As is derived in Sec. \ref{sec:aa-quadratic-speedup}, we see that the sample complexity can be reduced to roughly $1/\epsilon$ with a coherent quantum matching algorithm ($\swap$+AE), by combining amplitude estimation and a binary search protocol, thus achieving a quadratic speedup.
In this section, we present concrete numerical data demonstrating this quadratic speedup.

\begin{figure}[ht!]
    \centering
    \includegraphics[width=9cm]{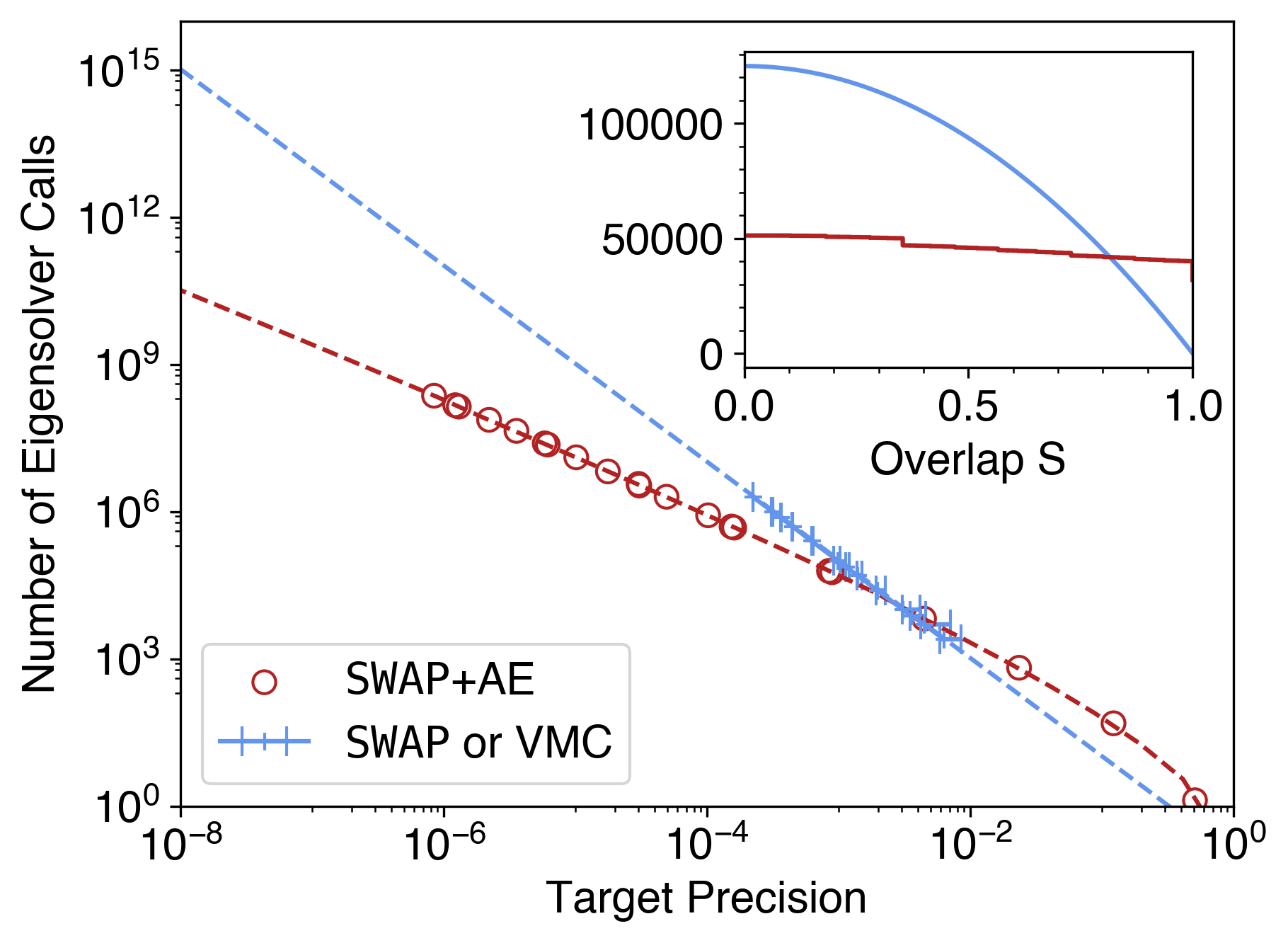}
    \caption{\label{fig:ampest_sqrt_speedup} Number of eigensolver calls required as a function of target precision at overlap $S=0.4$, comparing $\swap$ or VMC (blue) and $\swap$+AE (red) estimation for the H$_8$ chain with STO-3G basis. The blue dashed line shows the number of samples (eigensolver calls) needed in $\swap$ test as derived in Eq. \eqref{eq:nsamp_swap}, while the red dashed line plots a more accurate version of Eq. \eqref{eq:nsamp_swap+bs} (Sec. \ref{app:quadratic-speedup} of SI) with red circles highlighting a few data points spanning low to high target precisions. The blue scatter points are the number of VMC eigensolver calls required to achieve the corresponding target precision on the 1-RDM overlap estimation for the same H$_8$ molecule. The inset plots the number of eigensolver calls as a function of the overlap $S$ for a fixed target precision $\epsilon = 0.001$. Note the crossover in both plots.
    }
\end{figure}

Fig. \ref{fig:ampest_sqrt_speedup} shows that for a single BE iteration, the required number of samples (eigensolver calls) on estimating the RDM overlap $S$ between two adjacent fragments as a function of the required precision on the overlap, comparing the $\swap$ test based quantum matching (blue) and the coherent overlap estimation combining the $\swap$ test and amplitude estimation ($\swap$+AE) (red). We can see that the required number of samples increases quadratically as the accuracy $\epsilon$ increases for the $\swap$ test based estimation. In contrast, the slope of the $\swap$+AE sample complexity is reduced to roughly half of the $\swap$ test, demonstrating the quadratic speedup. 

To compare the classical VMC sampling convergence with the quantum overlap estimation method, we also overlay the number of VMC eigensolver calls (blue marks) versus target precision on estimating the overlap on top of the $\swap$ test sampling complexity for the same H$_8$ molecule. The general agreement between the VMC eigensolver calls and the derived $\swap$ test eigensolver calls highlights the similarity of a classical stochastic electronic structure methods and a quantum incoherent matching algorithm in terms of blackbox sampling complexity, echoing the idea of treating quantum computers as coherent sampling machines. It is worthwhile noting that this quadratic speedup is only advantageous in the high precision (small $\epsilon$) limit, as is evident from the existence of a crossing point in Fig. \ref{fig:ampest_sqrt_speedup} (between $10^{-4}$ and $10^{-2}$), which defines a critical $\epsilon^*$. For $\epsilon < \epsilon^*$, $\swap$+AE is favored whereas the $\swap$ test wins when $\epsilon > \epsilon^*$.

Moreover, in addition to the dependence on estimation accuracy $\epsilon$, the sampling complexity also depends on the value of the overlap $S$. The inset of Fig. \ref{fig:ampest_sqrt_speedup} compares the number of eigensolver calls using $\mathtt{SWAP}$ (blue) and the $\swap$+AE estimation (red) for estimating the overlap during quantum matching. In more detail, the sample complexity for the $\swap$ test decreases quadratically as the overlap $S$ approaches 1 (Eq. \eqref{eq:nsamp_swap}). As a comparison, the $\swap$+AE stays roughly a constant for the coherent quantum matching (\eqref{eq:nsamp_swap+bs}), because the amplitude amplification process used in the present work is agnostic to the value of the amplitude (overlap $S$), i.e., oblivious amplitude amplification \cite{yoder2014fixed,berry2014exponential}. The slight drop in sample complexity in the $\swap$+AE approach (red line, inset of Fig. \ref{fig:ampest_sqrt_speedup}) is due to the discrete bit representation of $S$ (Sec. \ref{app:binary-search} of SI). The different scaling on $S$ between these two algorithms leads to a crossover of the sampling complexity at roughly $S=0.8$ for a target precision of $\epsilon = 0.001$. This crossover suggests again that the plain $\swap$ test is advantageous for a large overlap, while amplitude estimation works better for small overlap $S$.

In addition, as mentioned in the previous section, as the bootstrap embedding iteration proceeds, the exponential convergence of the density mismatch (overlap $S$) suggests the need for an exponentially increasing accuracy $\epsilon$ on the overlap estimation. This further means the number of samples per iteration in the $\swap$ test should increases exponentially as the the number of iterations. 
Similarly, $\swap$+AE achieves a square-root speedup in the total sample numbers (remains exponential). We note that there may exist ways of sampling the overlap in the current BE iteration \emph{normalized} by the previous BE iteration to accelerate this requirement on a large number of samples, which we leave for future investigation.

\section{Conclusion and Outlook}
\label{sec:conclusion}

In conclusion, we have developed a general quantum bootstrap embedding method to find the ground state of large electronic structure problems on a quantum computer by taking advantage of quantum algorithms. We formulated the original electronic structure problem as a optimization problem using a quadratic penalty to impose matching condition of adjacent fragments. A coherent quantum matching algorithm based on the $\mathtt{SWAP}$ test achieves efficient matching with an exponential sampling advantage compared to naive RDM tomography. By estimating the amplitude that encodes the overlap information combing an amplitude amplification and binary search protocol, an additional quadratic speedup is achieved. In addition, an adaptive sampling scheme is used based on previous overlap information and the desired target accuracy to improve the sampling efficiency.

We demonstrate the performance of the QBE algorithm using a linear hydrogen molecule under minimal basis. Our QBE algorithm is shown to achieve  exponential convergence in density mismatch and energy error similar to classical bootstrap embedding. However, instead of the exponential cost of an exact classical solver (full configuration interaction), quantum eigensolvers such as quantum phase estimation can solve the fragment electronic structure exactly without incurring the exponential cost. Approximate quantum eigensolvers (QES) are likely to achieve exponential speedup compared to FCI. However, such exponential speedup depends on detailed implementation and the easy of input state preparation.



\begin{figure}[ht!]
    \centering
    \includegraphics[width=9cm]{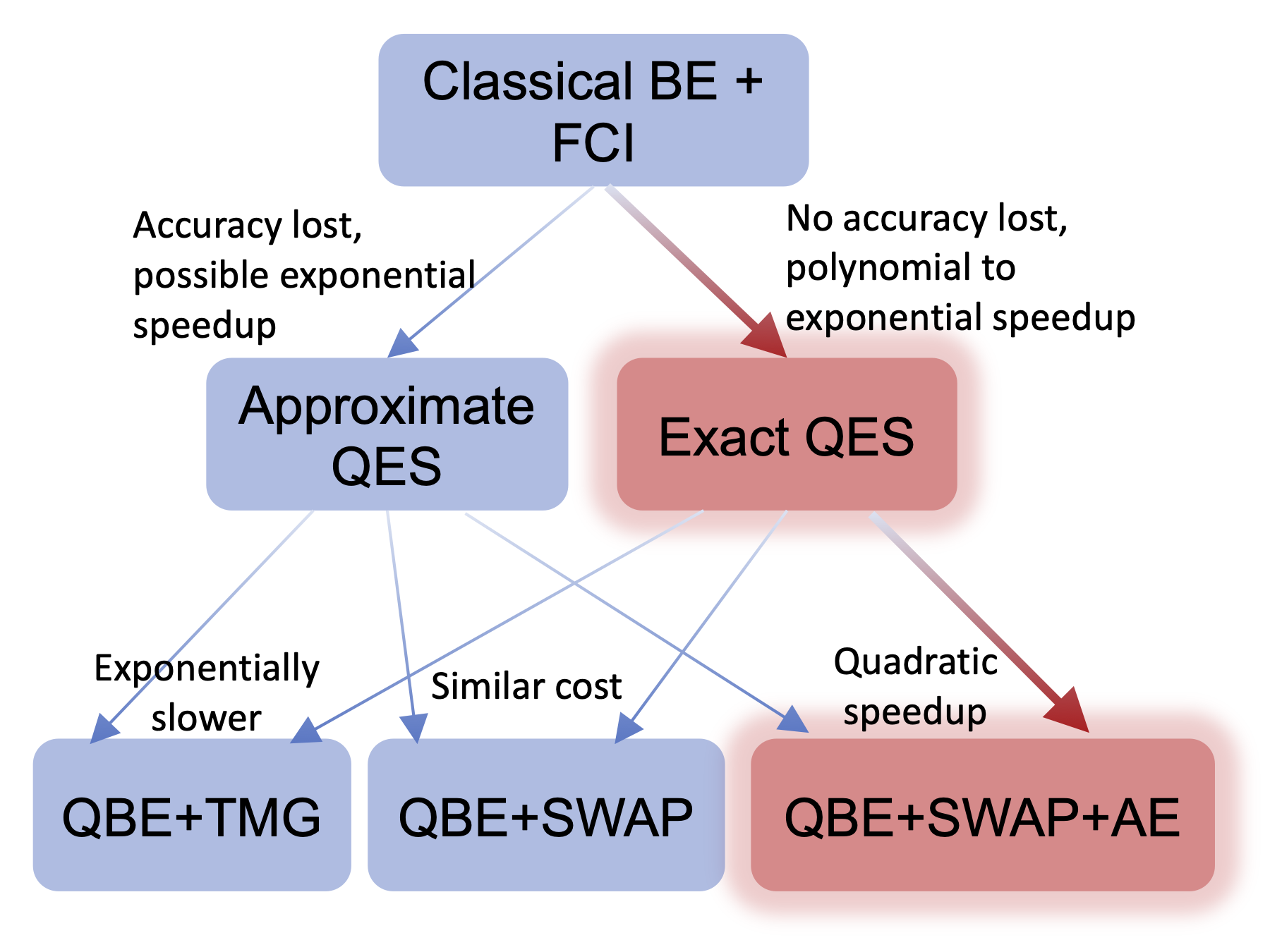}
    \caption{\label{fig:flow-chart} Summary of different choices of quantum eigensolvers (QES) and matching algorithms discussed in the present work, with speedup and cost labeled on each arrow accordingly. Overall, the best algorithm (QBE+SWAP+AE with exact QES) is highlighted in red. Note that approximate QES are likely to achieve exponential speedup as compared to classical FCI solver. It is however not guaranteed and depends on specific implementation and the ease of input state preparation. We therefore use "possible exponential speedup" for it. 
    }
\end{figure}

We have also compared sampling advantage of different versions of quantum matching algorithms over classical BE+VMC 1-RDM matching for achieving the same accuracy, 
where QBE+TMG (full RDM matching) is potentially exponentially slower than classical BE+VMC (1-RDM matching) because the exponentially large number of full RDM elements to estimate (Sec. \ref{sec:alg-eigensolver} and \ref{sec:results-sample-fragment-size}). QBE+$\swap$+AE achieves quadratic speedup as compared to classical BE+VMC and QBE+$\swap$ (Sec. \ref{sec:aa-quadratic-speedup} and \ref{sec:results-sample-quadratic-speedup}). Different choices of quantum eigensolvers and matching algorithms are summarized in the flow chart in Fig. \ref{fig:flow-chart}, where accuracy and speedups are labeled for each method.

While we have made progress toward solving electronic structure problems employing quantum resources in bootstrap embedding, there are several open questions to explore in the future. 
One immediate task is to perform more thorough benchmark comparing different versions of QBE and classical BE algorithms in terms of both speedup and accuracy quantitatively. Beyond benchmark, at the algorithmic level,
it is important to reconstruct \cite{qi2021emergent,nusspickel2022effective} the total system density matrices from subsystem ones in order to compute observables other than the energy. Ideally, quantum algorithms that can perform the reconstruction process would be desired.
Moreover, we have established how the bootstrap embedding potential can affect the system energy including the excited states in Eq.~\eqref{quadratic-gradient}. Future works on developing a QBE algorithm targeting excited states  \cite{mitra2021excited} or finite temperature electronic structures \cite{zhang1999finite,liu2018ab,sun2020finite} would be of great interest. Alternative constraint optimization methods such as the augmented Lagrangian method can also be explored to achieve potentially better convergence \cite{faulstich2022pure}.

In addition, the idea of quantum matching proposed in the present work could also be exploited further in other embedding theories to harness quantum computers and resources, including but not limited to embedding schemes based on wave functions, density matrices, and Green's functions \cite{sun2016quantum}. In these contexts, it is likely that more sophisticated quantum primitives and algorithms could accomplish quantum matching more efficiently  than the simple $\swap$ test we employ.  For example, it is possible that higher order matching, or matching of derivatives, could be accomplished quantum-mechanically, thus side-stepping sampling noise.

More broadly, these quantum embedding theories and algorithms enabled by quantum computation resources open new possibilities in chemistry, physics, and quantum information. In the near term, molecules with more complex valence electronic structures such as polyacetylene or polyacene chains beyond minimal basis can be treated with QBE on current noisy quantum computers with a few hundred qubits. In the longer term, large molecular systems in catalysis \cite{freeze2019search,zhou2012introduction}
and protein-ligand binding complexes \cite{warshel2014multiscale,proppe2020bioinspiration} likely can be simulated at a much higher accuracy by combining state-of-the-art quantum and classical computational resources in embedding properly. In condensed matter and material science, quantum bootstrap embedding may be adapted to periodic systems \cite{pham2019periodic,rusakov2018self,chibani2016self} for quantum material design \cite{head2020quantum} and probing phase diagrams of various lattice models \cite{qin2022hubbard} close to the thermodynamic limit.

Finally, from a viewpoint of quantum information, the concept of embedding is closely related to entanglement. Understanding the connection between the performance of quantum embedding algorithms and fragment-bath entanglement entropy may provide a general way to describe and understand the complexity of chemical and physical problems from a quantum information perspective \cite{ding2020concept,ding2020correlation,wilde2013quantum}. Current quantum computers are small -- we believe our quantum bootstrap embedding method provides a general strategy to use multiple small quantum machines to solve large problems in chemistry and beyond \cite{harrow2020small,song2021tangible}. We look forward to future development in these directions.

\begin{acknowledgement}
YL thanks Di Luo, Minh Tran, and Daniel Ranard for helpful discussions. The work on analysis and numerical simulation was supported by the U.S. Department of Energy, Office of Science, National Quantum Information Science Research Centers, Co-Design Center for Quantum Advantage, under contract number DE-SC0012704.  The conceptual algorithm development was supported in part by NTT Research.
\end{acknowledgement}

\begin{suppinfo}

Additional theoretical and numerical details.

\end{suppinfo}

\bibliography{ref}

\providecommand{\latin}[1]{#1}
\makeatletter
\providecommand{\doi}
  {\begingroup\let\do\@makeother\dospecials
  \catcode`\{=1 \catcode`\}=2 \doi@aux}
\providecommand{\doi@aux}[1]{\endgroup\texttt{#1}}
\makeatother
\providecommand*\mcitethebibliography{\thebibliography}
\csname @ifundefined\endcsname{endmcitethebibliography}
  {\let\endmcitethebibliography\endthebibliography}{}
\begin{mcitethebibliography}{110}
\providecommand*\natexlab[1]{#1}
\providecommand*\mciteSetBstSublistMode[1]{}
\providecommand*\mciteSetBstMaxWidthForm[2]{}
\providecommand*\mciteBstWouldAddEndPuncttrue
  {\def\EndOfBibitem{\unskip.}}
\providecommand*\mciteBstWouldAddEndPunctfalse
  {\let\EndOfBibitem\relax}
\providecommand*\mciteSetBstMidEndSepPunct[3]{}
\providecommand*\mciteSetBstSublistLabelBeginEnd[3]{}
\providecommand*\EndOfBibitem{}
\mciteSetBstSublistMode{f}
\mciteSetBstMaxWidthForm{subitem}{(\alph{mcitesubitemcount})}
\mciteSetBstSublistLabelBeginEnd
  {\mcitemaxwidthsubitemform\space}
  {\relax}
  {\relax}

\bibitem[Fukui \latin{et~al.}(1952)Fukui, Yonezawa, and
  Shingu]{fukui1952molecular}
Fukui,~K.; Yonezawa,~T.; Shingu,~H. A molecular orbital theory of reactivity in
  aromatic hydrocarbons. \emph{The Journal of Chemical Physics} \textbf{1952},
  \emph{20}, 722--725\relax
\mciteBstWouldAddEndPuncttrue
\mciteSetBstMidEndSepPunct{\mcitedefaultmidpunct}
{\mcitedefaultendpunct}{\mcitedefaultseppunct}\relax
\EndOfBibitem
\bibitem[Parr and Yang(1984)Parr, and Yang]{parr1984density}
Parr,~R.~G.; Yang,~W. Density functional approach to the frontier-electron
  theory of chemical reactivity. \emph{Journal of the American Chemical
  Society} \textbf{1984}, \emph{106}, 4049--4050\relax
\mciteBstWouldAddEndPuncttrue
\mciteSetBstMidEndSepPunct{\mcitedefaultmidpunct}
{\mcitedefaultendpunct}{\mcitedefaultseppunct}\relax
\EndOfBibitem
\bibitem[Greeley \latin{et~al.}(2002)Greeley, N{\o}rskov, and
  Mavrikakis]{greeley2002electronic}
Greeley,~J.; N{\o}rskov,~J.~K.; Mavrikakis,~M. Electronic structure and
  catalysis on metal surfaces. \emph{Annual Review of Physical Chemistry}
  \textbf{2002}, \emph{53}, 319--348\relax
\mciteBstWouldAddEndPuncttrue
\mciteSetBstMidEndSepPunct{\mcitedefaultmidpunct}
{\mcitedefaultendpunct}{\mcitedefaultseppunct}\relax
\EndOfBibitem
\bibitem[LeBlanc \latin{et~al.}(2015)LeBlanc, Antipov, Becca, Bulik, Chan,
  Chung, Deng, Ferrero, Henderson, Jim\'enez-Hoyos, Kozik, Liu, Millis,
  Prokof'ev, Qin, Scuseria, Shi, Svistunov, Tocchio, Tupitsyn, White, Zhang,
  Zheng, Zhu, and Gull]{leblanc2015solutions}
LeBlanc,~J. P.~F.; Antipov,~A.~E.; Becca,~F.; Bulik,~I.~W.; Chan,~G. K.-L.;
  Chung,~C.-M.; Deng,~Y.; Ferrero,~M.; Henderson,~T.~M.;
  Jim\'enez-Hoyos,~C.~A.; Kozik,~E.; Liu,~X.-W.; Millis,~A.~J.;
  Prokof'ev,~N.~V.; Qin,~M.; Scuseria,~G.~E.; Shi,~H.; Svistunov,~B.~V.;
  Tocchio,~L.~F.; Tupitsyn,~I.~S.; White,~S.~R.; Zhang,~S.; Zheng,~B.-X.;
  Zhu,~Z.; Gull,~E. Solutions of the two-dimensional Hubbard model: benchmarks
  and results from a wide range of numerical algorithms. \emph{Physical Review
  X} \textbf{2015}, \emph{5}, 041041\relax
\mciteBstWouldAddEndPuncttrue
\mciteSetBstMidEndSepPunct{\mcitedefaultmidpunct}
{\mcitedefaultendpunct}{\mcitedefaultseppunct}\relax
\EndOfBibitem
\bibitem[Zheng and Wagner(2015)Zheng, and Wagner]{zheng2015computation}
Zheng,~H.; Wagner,~L.~K. Computation of the correlated metal-insulator
  transition in vanadium dioxide from first principles. \emph{Physical Review
  Letters} \textbf{2015}, \emph{114}, 176401\relax
\mciteBstWouldAddEndPuncttrue
\mciteSetBstMidEndSepPunct{\mcitedefaultmidpunct}
{\mcitedefaultendpunct}{\mcitedefaultseppunct}\relax
\EndOfBibitem
\bibitem[Kotliar \latin{et~al.}(2006)Kotliar, Savrasov, Haule, Oudovenko,
  Parcollet, and Marianetti]{kotliar2006electronic}
Kotliar,~G.; Savrasov,~S.~Y.; Haule,~K.; Oudovenko,~V.~S.; Parcollet,~O.;
  Marianetti,~C. Electronic structure calculations with dynamical mean-field
  theory. \emph{Reviews of Modern Physics} \textbf{2006}, \emph{78}, 865\relax
\mciteBstWouldAddEndPuncttrue
\mciteSetBstMidEndSepPunct{\mcitedefaultmidpunct}
{\mcitedefaultendpunct}{\mcitedefaultseppunct}\relax
\EndOfBibitem
\bibitem[Gordon \latin{et~al.}(2012)Gordon, Fedorov, Pruitt, and
  Slipchenko]{gordon2012fragmentation}
Gordon,~M.~S.; Fedorov,~D.~G.; Pruitt,~S.~R.; Slipchenko,~L.~V. Fragmentation
  methods: A route to accurate calculations on large systems. \emph{Chemical
  Reviews} \textbf{2012}, \emph{112}, 632--672\relax
\mciteBstWouldAddEndPuncttrue
\mciteSetBstMidEndSepPunct{\mcitedefaultmidpunct}
{\mcitedefaultendpunct}{\mcitedefaultseppunct}\relax
\EndOfBibitem
\bibitem[Jones \latin{et~al.}(2020)Jones, Mosquera, Schatz, and
  Ratner]{jones2020embedding}
Jones,~L.~O.; Mosquera,~M.~A.; Schatz,~G.~C.; Ratner,~M.~A. Embedding methods
  for quantum chemistry: applications from materials to life sciences.
  \emph{Journal of the American Chemical Society} \textbf{2020}, \emph{142},
  3281--3295\relax
\mciteBstWouldAddEndPuncttrue
\mciteSetBstMidEndSepPunct{\mcitedefaultmidpunct}
{\mcitedefaultendpunct}{\mcitedefaultseppunct}\relax
\EndOfBibitem
\bibitem[Sun and Chan(2016)Sun, and Chan]{sun2016quantum}
Sun,~Q.; Chan,~G. K.-L. Quantum embedding theories. \emph{Accounts of Chemical
  Research} \textbf{2016}, \emph{49}, 2705--2712\relax
\mciteBstWouldAddEndPuncttrue
\mciteSetBstMidEndSepPunct{\mcitedefaultmidpunct}
{\mcitedefaultendpunct}{\mcitedefaultseppunct}\relax
\EndOfBibitem
\bibitem[Wesolowski \latin{et~al.}(2015)Wesolowski, Shedge, and
  Zhou]{wesolowski2015frozen}
Wesolowski,~T.~A.; Shedge,~S.; Zhou,~X. Frozen-density embedding strategy for
  multilevel simulations of electronic structure. \emph{Chemical Reviews}
  \textbf{2015}, \emph{115}, 5891--5928\relax
\mciteBstWouldAddEndPuncttrue
\mciteSetBstMidEndSepPunct{\mcitedefaultmidpunct}
{\mcitedefaultendpunct}{\mcitedefaultseppunct}\relax
\EndOfBibitem
\bibitem[Libisch \latin{et~al.}(2014)Libisch, Huang, and
  Carter]{libisch2014embedded}
Libisch,~F.; Huang,~C.; Carter,~E.~A. Embedded correlated wavefunction schemes:
  Theory and applications. \emph{Accounts of Chemical Research} \textbf{2014},
  \emph{47}, 2768--2775\relax
\mciteBstWouldAddEndPuncttrue
\mciteSetBstMidEndSepPunct{\mcitedefaultmidpunct}
{\mcitedefaultendpunct}{\mcitedefaultseppunct}\relax
\EndOfBibitem
\bibitem[Knizia and Chan(2012)Knizia, and Chan]{knizia2012density}
Knizia,~G.; Chan,~G. K.-L. Density matrix embedding: A simple alternative to
  dynamical mean-field theory. \emph{Physical Review Letters} \textbf{2012},
  \emph{109}, 186404\relax
\mciteBstWouldAddEndPuncttrue
\mciteSetBstMidEndSepPunct{\mcitedefaultmidpunct}
{\mcitedefaultendpunct}{\mcitedefaultseppunct}\relax
\EndOfBibitem
\bibitem[Knizia and Chan(2013)Knizia, and Chan]{knizia2013density}
Knizia,~G.; Chan,~G. K.-L. Density matrix embedding: A strong-coupling quantum
  embedding theory. \emph{Journal of Chemical Theory and Computation}
  \textbf{2013}, \emph{9}, 1428--1432\relax
\mciteBstWouldAddEndPuncttrue
\mciteSetBstMidEndSepPunct{\mcitedefaultmidpunct}
{\mcitedefaultendpunct}{\mcitedefaultseppunct}\relax
\EndOfBibitem
\bibitem[Wouters \latin{et~al.}(2016)Wouters, Jim{\'e}nez-Hoyos, Sun, and
  Chan]{wouters2016practical}
Wouters,~S.; Jim{\'e}nez-Hoyos,~C.~A.; Sun,~Q.; Chan,~G. K.-L. A practical
  guide to density matrix embedding theory in quantum chemistry. \emph{Journal
  of Chemical Theory and Computation} \textbf{2016}, \emph{12},
  2706--2719\relax
\mciteBstWouldAddEndPuncttrue
\mciteSetBstMidEndSepPunct{\mcitedefaultmidpunct}
{\mcitedefaultendpunct}{\mcitedefaultseppunct}\relax
\EndOfBibitem
\bibitem[Wouters \latin{et~al.}(2017)Wouters, A.~Jim{\'e}nez-Hoyos, and
  KL~Chan]{wouters2017five}
Wouters,~S.; A.~Jim{\'e}nez-Hoyos,~C.; KL~Chan,~G. Five years of density matrix
  embedding theory. \emph{Fragmentation: toward accurate calculations on
  complex molecular systems} \textbf{2017}, 227--243\relax
\mciteBstWouldAddEndPuncttrue
\mciteSetBstMidEndSepPunct{\mcitedefaultmidpunct}
{\mcitedefaultendpunct}{\mcitedefaultseppunct}\relax
\EndOfBibitem
\bibitem[Faulstich \latin{et~al.}(2022)Faulstich, Kim, Cui, Wen, Kin-Lic~Chan,
  and Lin]{faulstich2022pure}
Faulstich,~F.~M.; Kim,~R.; Cui,~Z.-H.; Wen,~Z.; Kin-Lic~Chan,~G.; Lin,~L. Pure
  State v-Representability of Density Matrix Embedding Theory. \emph{Journal of
  Chemical Theory and Computation} \textbf{2022}, \emph{18}, 851--864\relax
\mciteBstWouldAddEndPuncttrue
\mciteSetBstMidEndSepPunct{\mcitedefaultmidpunct}
{\mcitedefaultendpunct}{\mcitedefaultseppunct}\relax
\EndOfBibitem
\bibitem[Hettler \latin{et~al.}(2000)Hettler, Mukherjee, Jarrell, and
  Krishnamurthy]{hettler2000dynamical}
Hettler,~M.; Mukherjee,~M.; Jarrell,~M.; Krishnamurthy,~H. Dynamical cluster
  approximation: Nonlocal dynamics of correlated electron systems.
  \emph{Physical Review B} \textbf{2000}, \emph{61}, 12739\relax
\mciteBstWouldAddEndPuncttrue
\mciteSetBstMidEndSepPunct{\mcitedefaultmidpunct}
{\mcitedefaultendpunct}{\mcitedefaultseppunct}\relax
\EndOfBibitem
\bibitem[Ma \latin{et~al.}(2021)Ma, Sheng, Govoni, and Galli]{ma2021quantum}
Ma,~H.; Sheng,~N.; Govoni,~M.; Galli,~G. Quantum embedding theory for strongly
  correlated states in materials. \emph{Journal of Chemical Theory and
  Computation} \textbf{2021}, \emph{17}, 2116--2125\relax
\mciteBstWouldAddEndPuncttrue
\mciteSetBstMidEndSepPunct{\mcitedefaultmidpunct}
{\mcitedefaultendpunct}{\mcitedefaultseppunct}\relax
\EndOfBibitem
\bibitem[Lan and Zgid(2017)Lan, and Zgid]{lan2017generalized}
Lan,~T.~N.; Zgid,~D. Generalized self-energy embedding theory. \emph{The
  Journal of Physical Chemistry Letters} \textbf{2017}, \emph{8},
  2200--2205\relax
\mciteBstWouldAddEndPuncttrue
\mciteSetBstMidEndSepPunct{\mcitedefaultmidpunct}
{\mcitedefaultendpunct}{\mcitedefaultseppunct}\relax
\EndOfBibitem
\bibitem[Rusakov \latin{et~al.}(2018)Rusakov, Iskakov, Tran, and
  Zgid]{rusakov2018self}
Rusakov,~A.~A.; Iskakov,~S.; Tran,~L.~N.; Zgid,~D. Self-energy embedding theory
  (SEET) for periodic systems. \emph{Journal of Chemical Theory and
  Computation} \textbf{2018}, \emph{15}, 229--240\relax
\mciteBstWouldAddEndPuncttrue
\mciteSetBstMidEndSepPunct{\mcitedefaultmidpunct}
{\mcitedefaultendpunct}{\mcitedefaultseppunct}\relax
\EndOfBibitem
\bibitem[Biermann \latin{et~al.}(2003)Biermann, Aryasetiawan, and
  Georges]{biermann2003first}
Biermann,~S.; Aryasetiawan,~F.; Georges,~A. First-Principles Approach to the
  Electronic Structure of Strongly Correlated Systems: Combining the G W
  Approximation and Dynamical Mean-Field Theory. \emph{Physical Review Letters}
  \textbf{2003}, \emph{90}, 086402\relax
\mciteBstWouldAddEndPuncttrue
\mciteSetBstMidEndSepPunct{\mcitedefaultmidpunct}
{\mcitedefaultendpunct}{\mcitedefaultseppunct}\relax
\EndOfBibitem
\bibitem[Welborn \latin{et~al.}(2016)Welborn, Tsuchimochi, and
  Van~Voorhis]{welborn2016bootstrap}
Welborn,~M.; Tsuchimochi,~T.; Van~Voorhis,~T. Bootstrap embedding: An
  internally consistent fragment-based method. \emph{The Journal of Chemical
  Physics} \textbf{2016}, \emph{145}, 074102\relax
\mciteBstWouldAddEndPuncttrue
\mciteSetBstMidEndSepPunct{\mcitedefaultmidpunct}
{\mcitedefaultendpunct}{\mcitedefaultseppunct}\relax
\EndOfBibitem
\bibitem[Ye \latin{et~al.}(2019)Ye, Ricke, Tran, and
  Van~Voorhis]{ye2019bootstrap}
Ye,~H.-Z.; Ricke,~N.~D.; Tran,~H.~K.; Van~Voorhis,~T. Bootstrap embedding for
  molecules. \emph{Journal of Chemical Theory and Computation} \textbf{2019},
  \emph{15}, 4497--4506\relax
\mciteBstWouldAddEndPuncttrue
\mciteSetBstMidEndSepPunct{\mcitedefaultmidpunct}
{\mcitedefaultendpunct}{\mcitedefaultseppunct}\relax
\EndOfBibitem
\bibitem[Ye \latin{et~al.}(2021)Ye, Tran, and Van~Voorhis]{ye2021accurate}
Ye,~H.-Z.; Tran,~H.~K.; Van~Voorhis,~T. Accurate Electronic Excitation Energies
  in Full-Valence Active Space via Bootstrap Embedding. \emph{Journal of
  Chemical Theory and Computation} \textbf{2021}, \emph{17}, 3335--3347\relax
\mciteBstWouldAddEndPuncttrue
\mciteSetBstMidEndSepPunct{\mcitedefaultmidpunct}
{\mcitedefaultendpunct}{\mcitedefaultseppunct}\relax
\EndOfBibitem
\bibitem[Zheng \latin{et~al.}(2017)Zheng, Chung, Corboz, Ehlers, Qin, Noack,
  Shi, White, Zhang, and Chan]{zheng2017stripe}
Zheng,~B.-X.; Chung,~C.-M.; Corboz,~P.; Ehlers,~G.; Qin,~M.-P.; Noack,~R.~M.;
  Shi,~H.; White,~S.~R.; Zhang,~S.; Chan,~G. K.-L. Stripe order in the
  underdoped region of the two-dimensional Hubbard model. \emph{Science}
  \textbf{2017}, \emph{358}, 1155--1160\relax
\mciteBstWouldAddEndPuncttrue
\mciteSetBstMidEndSepPunct{\mcitedefaultmidpunct}
{\mcitedefaultendpunct}{\mcitedefaultseppunct}\relax
\EndOfBibitem
\bibitem[Zhu \latin{et~al.}(2019)Zhu, Jim{\'e}nez-Hoyos, McClain, Berkelbach,
  and Chan]{zhu2019coupled}
Zhu,~T.; Jim{\'e}nez-Hoyos,~C.~A.; McClain,~J.; Berkelbach,~T.~C.; Chan,~G.
  K.-L. Coupled-cluster impurity solvers for dynamical mean-field theory.
  \emph{Physical Review B} \textbf{2019}, \emph{100}, 115154\relax
\mciteBstWouldAddEndPuncttrue
\mciteSetBstMidEndSepPunct{\mcitedefaultmidpunct}
{\mcitedefaultendpunct}{\mcitedefaultseppunct}\relax
\EndOfBibitem
\bibitem[Shee and Zgid(2019)Shee, and Zgid]{shee2019coupled}
Shee,~A.; Zgid,~D. Coupled cluster as an impurity solver for Green’s function
  embedding methods. \emph{Journal of Chemical Theory and Computation}
  \textbf{2019}, \emph{15}, 6010--6024\relax
\mciteBstWouldAddEndPuncttrue
\mciteSetBstMidEndSepPunct{\mcitedefaultmidpunct}
{\mcitedefaultendpunct}{\mcitedefaultseppunct}\relax
\EndOfBibitem
\bibitem[Lau \latin{et~al.}(2021)Lau, Knizia, and Berkelbach]{lau2021regional}
Lau,~B.~T.; Knizia,~G.; Berkelbach,~T.~C. Regional embedding enables high-level
  quantum chemistry for surface science. \emph{The Journal of Physical
  Chemistry Letters} \textbf{2021}, \emph{12}, 1104--1109\relax
\mciteBstWouldAddEndPuncttrue
\mciteSetBstMidEndSepPunct{\mcitedefaultmidpunct}
{\mcitedefaultendpunct}{\mcitedefaultseppunct}\relax
\EndOfBibitem
\bibitem[Bauer \latin{et~al.}(2020)Bauer, Bravyi, Motta, and
  Chan]{bauer2020quantum}
Bauer,~B.; Bravyi,~S.; Motta,~M.; Chan,~G. K.-L. Quantum algorithms for quantum
  chemistry and quantum materials science. \emph{Chemical Reviews}
  \textbf{2020}, \emph{120}, 12685--12717\relax
\mciteBstWouldAddEndPuncttrue
\mciteSetBstMidEndSepPunct{\mcitedefaultmidpunct}
{\mcitedefaultendpunct}{\mcitedefaultseppunct}\relax
\EndOfBibitem
\bibitem[Lee \latin{et~al.}(2022)Lee, Lee, Zhai, Tong, Dalzell, Kumar, Helms,
  Gray, Cui, Liu, Kastoryano, Babbush, Preskill, Reichman, Campbell, Valeev,
  Lin, and Chan]{lee2022there}
Lee,~S.; Lee,~J.; Zhai,~H.; Tong,~Y.; Dalzell,~A.~M.; Kumar,~A.; Helms,~P.;
  Gray,~J.; Cui,~Z.-H.; Liu,~W.; Kastoryano,~M.; Babbush,~R.; Preskill,~J.;
  Reichman,~D.~R.; Campbell,~E.~T.; Valeev,~E.~F.; Lin,~L.; Chan,~G. K.-L. Is
  there evidence for exponential quantum advantage in quantum chemistry?
  \emph{arXiv preprint arXiv:2208.02199} \textbf{2022}, \relax
\mciteBstWouldAddEndPunctfalse
\mciteSetBstMidEndSepPunct{\mcitedefaultmidpunct}
{}{\mcitedefaultseppunct}\relax
\EndOfBibitem
\bibitem[Abrams and Lloyd(1999)Abrams, and Lloyd]{abrams1999quantum}
Abrams,~D.~S.; Lloyd,~S. Quantum algorithm providing exponential speed increase
  for finding eigenvalues and eigenvectors. \emph{Physical Review Letters}
  \textbf{1999}, \emph{83}, 5162\relax
\mciteBstWouldAddEndPuncttrue
\mciteSetBstMidEndSepPunct{\mcitedefaultmidpunct}
{\mcitedefaultendpunct}{\mcitedefaultseppunct}\relax
\EndOfBibitem
\bibitem[Aspuru-Guzik \latin{et~al.}(2005)Aspuru-Guzik, Dutoi, Love, and
  Head-Gordon]{aspuru2005simulated}
Aspuru-Guzik,~A.; Dutoi,~A.~D.; Love,~P.~J.; Head-Gordon,~M. Simulated quantum
  computation of molecular energies. \emph{Science} \textbf{2005}, \emph{309},
  1704--1707\relax
\mciteBstWouldAddEndPuncttrue
\mciteSetBstMidEndSepPunct{\mcitedefaultmidpunct}
{\mcitedefaultendpunct}{\mcitedefaultseppunct}\relax
\EndOfBibitem
\bibitem[Peruzzo \latin{et~al.}(2014)Peruzzo, McClean, Shadbolt, Yung, Zhou,
  Love, Aspuru-Guzik, and O'Brien]{Peruzzo2014variational}
Peruzzo,~A.; McClean,~J.; Shadbolt,~P.; Yung,~M.-H.; Zhou,~X.-Q.; Love,~P.~J.;
  Aspuru-Guzik,~A.; O'Brien,~J.~L. A variational eigenvalue solver on a
  photonic quantum processor. \emph{Nature Communications} \textbf{2014},
  \emph{5}\relax
\mciteBstWouldAddEndPuncttrue
\mciteSetBstMidEndSepPunct{\mcitedefaultmidpunct}
{\mcitedefaultendpunct}{\mcitedefaultseppunct}\relax
\EndOfBibitem
\bibitem[Tilly \latin{et~al.}(2022)Tilly, Chen, Cao, Picozzi, Setia, Li, Grant,
  Wossnig, Rungger, Booth, and Tennyson]{tilly2022variational}
Tilly,~J.; Chen,~H.; Cao,~S.; Picozzi,~D.; Setia,~K.; Li,~Y.; Grant,~E.;
  Wossnig,~L.; Rungger,~I.; Booth,~G.~H.; Tennyson,~J. The variational quantum
  eigensolver: a review of methods and best practices. \emph{Physics Reports}
  \textbf{2022}, \emph{986}, 1--128\relax
\mciteBstWouldAddEndPuncttrue
\mciteSetBstMidEndSepPunct{\mcitedefaultmidpunct}
{\mcitedefaultendpunct}{\mcitedefaultseppunct}\relax
\EndOfBibitem
\bibitem[Wang \latin{et~al.}(2019)Wang, Higgott, and
  Brierley]{wang2019accelerated}
Wang,~D.; Higgott,~O.; Brierley,~S. Accelerated variational quantum
  eigensolver. \emph{Physical Review Letters} \textbf{2019}, \emph{122},
  140504\relax
\mciteBstWouldAddEndPuncttrue
\mciteSetBstMidEndSepPunct{\mcitedefaultmidpunct}
{\mcitedefaultendpunct}{\mcitedefaultseppunct}\relax
\EndOfBibitem
\bibitem[Grimsley \latin{et~al.}(2019)Grimsley, Economou, Barnes, and
  Mayhall]{grimsley2019adaptive}
Grimsley,~H.~R.; Economou,~S.~E.; Barnes,~E.; Mayhall,~N.~J. An adaptive
  variational algorithm for exact molecular simulations on a quantum computer.
  \emph{Nature Communications} \textbf{2019}, \emph{10}, 1--9\relax
\mciteBstWouldAddEndPuncttrue
\mciteSetBstMidEndSepPunct{\mcitedefaultmidpunct}
{\mcitedefaultendpunct}{\mcitedefaultseppunct}\relax
\EndOfBibitem
\bibitem[Grimsley \latin{et~al.}(2022)Grimsley, Barron, Barnes, Economou, and
  Mayhall]{grimsley2022adaptvqe}
Grimsley,~H.~R.; Barron,~G.~S.; Barnes,~E.; Economou,~S.~E.; Mayhall,~N.~J.
  ADAPT-VQE is insensitive to rough parameter landscapes and barren plateaus.
  \emph{arXiv preprint arXiv:2204.07179} \textbf{2022}, \relax
\mciteBstWouldAddEndPunctfalse
\mciteSetBstMidEndSepPunct{\mcitedefaultmidpunct}
{}{\mcitedefaultseppunct}\relax
\EndOfBibitem
\bibitem[Cramer \latin{et~al.}(2010)Cramer, Plenio, Flammia, Somma, Gross,
  Bartlett, Landon-Cardinal, Poulin, and Liu]{cramer2010efficient}
Cramer,~M.; Plenio,~M.~B.; Flammia,~S.~T.; Somma,~R.; Gross,~D.;
  Bartlett,~S.~D.; Landon-Cardinal,~O.; Poulin,~D.; Liu,~Y.-K. Efficient
  quantum state tomography. \emph{Nature Communications} \textbf{2010},
  \emph{1}, 1--7\relax
\mciteBstWouldAddEndPuncttrue
\mciteSetBstMidEndSepPunct{\mcitedefaultmidpunct}
{\mcitedefaultendpunct}{\mcitedefaultseppunct}\relax
\EndOfBibitem
\bibitem[Zhao \latin{et~al.}(2021)Zhao, Rubin, and Miyake]{zhao2021fermionic}
Zhao,~A.; Rubin,~N.~C.; Miyake,~A. Fermionic Partial Tomography via Classical
  Shadows. \emph{Physical Review Letters} \textbf{2021}, \emph{127}\relax
\mciteBstWouldAddEndPuncttrue
\mciteSetBstMidEndSepPunct{\mcitedefaultmidpunct}
{\mcitedefaultendpunct}{\mcitedefaultseppunct}\relax
\EndOfBibitem
\bibitem[Otten \latin{et~al.}(2022)Otten, Hermes, Pandharkar, Alexeev, Gray,
  and Gagliardi]{otten2022localized}
Otten,~M.; Hermes,~M.~R.; Pandharkar,~R.; Alexeev,~Y.; Gray,~S.~K.;
  Gagliardi,~L. Localized {Quantum} {Chemistry} on {Quantum} {Computers}.
  \emph{Journal of Chemical Theory and Computation} \textbf{2022}, \emph{18},
  7205--7217\relax
\mciteBstWouldAddEndPuncttrue
\mciteSetBstMidEndSepPunct{\mcitedefaultmidpunct}
{\mcitedefaultendpunct}{\mcitedefaultseppunct}\relax
\EndOfBibitem
\bibitem[Vorwerk \latin{et~al.}(2022)Vorwerk, Sheng, Govoni, Huang, and
  Galli]{vorwerk2022quantum}
Vorwerk,~C.; Sheng,~N.; Govoni,~M.; Huang,~B.; Galli,~G. Quantum embedding
  theories to simulate condensed systems on quantum computers. \emph{Nature
  Computational Science} \textbf{2022}, \emph{2}, 424--432\relax
\mciteBstWouldAddEndPuncttrue
\mciteSetBstMidEndSepPunct{\mcitedefaultmidpunct}
{\mcitedefaultendpunct}{\mcitedefaultseppunct}\relax
\EndOfBibitem
\bibitem[Mineh and Montanaro(2022)Mineh, and Montanaro]{mineh2022prb}
Mineh,~L.; Montanaro,~A. Solving the Hubbard model using density matrix
  embedding theory and the variational quantum eigensolver. \emph{Physical
  Review B} \textbf{2022}, \emph{105}, 125117\relax
\mciteBstWouldAddEndPuncttrue
\mciteSetBstMidEndSepPunct{\mcitedefaultmidpunct}
{\mcitedefaultendpunct}{\mcitedefaultseppunct}\relax
\EndOfBibitem
\bibitem[Li \latin{et~al.}(2022)Li, Huang, Cao, Huang, Shuai, Sun, Sun, Yuan,
  and Lv]{li2022toward}
Li,~W.; Huang,~Z.; Cao,~C.; Huang,~Y.; Shuai,~Z.; Sun,~X.; Sun,~J.; Yuan,~X.;
  Lv,~D. Toward practical quantum embedding simulation of realistic chemical
  systems on near-term quantum computers. \emph{Chemical Science}
  \textbf{2022}, \emph{13}, 8953--8962\relax
\mciteBstWouldAddEndPuncttrue
\mciteSetBstMidEndSepPunct{\mcitedefaultmidpunct}
{\mcitedefaultendpunct}{\mcitedefaultseppunct}\relax
\EndOfBibitem
\bibitem[Ma \latin{et~al.}(2020)Ma, Govoni, and Galli]{ma2020quantum}
Ma,~H.; Govoni,~M.; Galli,~G. Quantum simulations of materials on near-term
  quantum computers. \emph{npj Computational Materials} \textbf{2020},
  \emph{6}, 1--8\relax
\mciteBstWouldAddEndPuncttrue
\mciteSetBstMidEndSepPunct{\mcitedefaultmidpunct}
{\mcitedefaultendpunct}{\mcitedefaultseppunct}\relax
\EndOfBibitem
\bibitem[Quantum \latin{et~al.}(2020)Quantum, Collaborators*†, Arute, Arya,
  Babbush, Bacon, Bardin, Barends, Boixo, Broughton, Buckley, Buell, Burkett,
  Bushnell, Chen, Chen, Chiaro, Collins, Courtney, Demura, Dunsworth, Farhi,
  Fowler, Foxen, Gidney, Giustina, Graff, Habegger, Harrigan, Ho, Hong, Huang,
  Huggins, Ioffe, Isakov, Jeffrey, Jiang, Jones, Kafri, Kechedzhi, Kelly, Kim,
  Klimov, Korotkov, Kostritsa, Landhuis, Laptev, Lindmark, Lucero, Martin,
  Martinis, McClean, McEwen, Megrant, Mi, Mohseni, Mruczkiewicz, Mutus, Naaman,
  Neeley, Neill, Neven, Niu, O'Brien, Ostby, Petukhov, Putterman, Quintana,
  Roushan, Rubin, Sank, Satzinger, Smelyanskiy, Strain, Sung, Szalay,
  Takeshita, Vainsencher, White, Wiebe, Yao, Yeh, and
  Zalcman]{google2020hartree}
Quantum,~G.~A.; Collaborators*†,; Arute,~F.; Arya,~K.; Babbush,~R.;
  Bacon,~D.; Bardin,~J.~C.; Barends,~R.; Boixo,~S.; Broughton,~M.;
  Buckley,~B.~B.; Buell,~D.~A.; Burkett,~B.; Bushnell,~N.; Chen,~Y.; Chen,~Z.;
  Chiaro,~B.; Collins,~R.; Courtney,~W.; Demura,~S.; Dunsworth,~A.; Farhi,~E.;
  Fowler,~A.; Foxen,~B.; Gidney,~C.; Giustina,~M.; Graff,~R.; Habegger,~S.;
  Harrigan,~M.~P.; Ho,~A.; Hong,~S.; Huang,~T.; Huggins,~W.~J.; Ioffe,~L.;
  Isakov,~S.~V.; Jeffrey,~E.; Jiang,~Z.; Jones,~C.; Kafri,~D.; Kechedzhi,~K.;
  Kelly,~J.; Kim,~S.; Klimov,~P.~V.; Korotkov,~A.; Kostritsa,~F.; Landhuis,~D.;
  Laptev,~P.; Lindmark,~M.; Lucero,~E.; Martin,~O.; Martinis,~J.~M.;
  McClean,~J.~R.; McEwen,~M.; Megrant,~A.; Mi,~X.; Mohseni,~M.;
  Mruczkiewicz,~W.; Mutus,~J.; Naaman,~O.; Neeley,~M.; Neill,~C.; Neven,~H.;
  Niu,~M.~Y.; O'Brien,~T.~E.; Ostby,~E.; Petukhov,~A.; Putterman,~H.;
  Quintana,~C.; Roushan,~P.; Rubin,~N.~C.; Sank,~D.; Satzinger,~K.~J.;
  Smelyanskiy,~V.; Strain,~D.; Sung,~K.~J.; Szalay,~M.; Takeshita,~T.~Y.;
  Vainsencher,~A.; White,~T.; Wiebe,~N.; Yao,~Z.~J.; Yeh,~P.; Zalcman,~A.
  Hartree-Fock on a superconducting qubit quantum computer. \emph{Science}
  \textbf{2020}, \emph{369}, 1084--1089\relax
\mciteBstWouldAddEndPuncttrue
\mciteSetBstMidEndSepPunct{\mcitedefaultmidpunct}
{\mcitedefaultendpunct}{\mcitedefaultseppunct}\relax
\EndOfBibitem
\bibitem[Barenco \latin{et~al.}(1997)Barenco, Berthiaume, Deutsch, Ekert,
  Jozsa, and Macchiavello]{barenco1997stabilization}
Barenco,~A.; Berthiaume,~A.; Deutsch,~D.; Ekert,~A.; Jozsa,~R.;
  Macchiavello,~C. Stabilization of Quantum Computations by Symmetrization.
  \emph{SIAM Journal on Computing} \textbf{1997}, \emph{26}, 1541--1557\relax
\mciteBstWouldAddEndPuncttrue
\mciteSetBstMidEndSepPunct{\mcitedefaultmidpunct}
{\mcitedefaultendpunct}{\mcitedefaultseppunct}\relax
\EndOfBibitem
\bibitem[Buhrman \latin{et~al.}(2001)Buhrman, Cleve, Watrous, and
  de~Wolf]{buhrman2001quantum}
Buhrman,~H.; Cleve,~R.; Watrous,~J.; de~Wolf,~R. Quantum Fingerprinting.
  \emph{Physical Review Letters} \textbf{2001}, \emph{87}, 167902\relax
\mciteBstWouldAddEndPuncttrue
\mciteSetBstMidEndSepPunct{\mcitedefaultmidpunct}
{\mcitedefaultendpunct}{\mcitedefaultseppunct}\relax
\EndOfBibitem
\bibitem[Brassard \latin{et~al.}(2002)Brassard, Hoyer, Mosca, and
  Tapp]{brassard2002quantum}
Brassard,~G.; Hoyer,~P.; Mosca,~M.; Tapp,~A. Quantum amplitude amplification
  and estimation. \emph{Contemporary Mathematics} \textbf{2002}, \emph{305},
  53--74\relax
\mciteBstWouldAddEndPuncttrue
\mciteSetBstMidEndSepPunct{\mcitedefaultmidpunct}
{\mcitedefaultendpunct}{\mcitedefaultseppunct}\relax
\EndOfBibitem
\bibitem[Martyn \latin{et~al.}(2021)Martyn, Rossi, Tan, and
  Chuang]{martyn2021grand}
Martyn,~J.~M.; Rossi,~Z.~M.; Tan,~A.~K.; Chuang,~I.~L. Grand unification of
  quantum algorithms. \emph{PRX Quantum} \textbf{2021}, \emph{2}, 040203\relax
\mciteBstWouldAddEndPuncttrue
\mciteSetBstMidEndSepPunct{\mcitedefaultmidpunct}
{\mcitedefaultendpunct}{\mcitedefaultseppunct}\relax
\EndOfBibitem
\bibitem[Ye \latin{et~al.}(2020)Ye, Tran, and Van~Voorhis]{ye2020bootstrap}
Ye,~H.-Z.; Tran,~H.~K.; Van~Voorhis,~T. Bootstrap embedding for large molecular
  systems. \emph{Journal of Chemical Theory and Computation} \textbf{2020},
  \emph{16}, 5035--5046\relax
\mciteBstWouldAddEndPuncttrue
\mciteSetBstMidEndSepPunct{\mcitedefaultmidpunct}
{\mcitedefaultendpunct}{\mcitedefaultseppunct}\relax
\EndOfBibitem
\bibitem[L{\"o}wdin(1950)]{lowdin1950non}
L{\"o}wdin,~P.-O. On the non-orthogonality problem connected with the use of
  atomic wave functions in the theory of molecules and crystals. \emph{The
  Journal of Chemical Physics} \textbf{1950}, \emph{18}, 365--375\relax
\mciteBstWouldAddEndPuncttrue
\mciteSetBstMidEndSepPunct{\mcitedefaultmidpunct}
{\mcitedefaultendpunct}{\mcitedefaultseppunct}\relax
\EndOfBibitem
\bibitem[Claudino and Mayhall(2019)Claudino, and
  Mayhall]{claudino2019automatic}
Claudino,~D.; Mayhall,~N.~J. Automatic partition of orbital spaces based on
  singular value decomposition in the context of embedding theories.
  \emph{Journal of Chemical Theory and Computation} \textbf{2019}, \emph{15},
  1053--1064\relax
\mciteBstWouldAddEndPuncttrue
\mciteSetBstMidEndSepPunct{\mcitedefaultmidpunct}
{\mcitedefaultendpunct}{\mcitedefaultseppunct}\relax
\EndOfBibitem
\bibitem[Waldrop \latin{et~al.}(2021)Waldrop, Windus, and
  Govind]{waldrop2021projector}
Waldrop,~J.~M.; Windus,~T.~L.; Govind,~N. Projector-based quantum embedding for
  molecular systems: An investigation of three partitioning approaches.
  \emph{The Journal of Physical Chemistry A} \textbf{2021}, \emph{125},
  6384--6393\relax
\mciteBstWouldAddEndPuncttrue
\mciteSetBstMidEndSepPunct{\mcitedefaultmidpunct}
{\mcitedefaultendpunct}{\mcitedefaultseppunct}\relax
\EndOfBibitem
\bibitem[Ye and Van~Voorhis(2019)Ye, and Van~Voorhis]{ye2019atom}
Ye,~H.-Z.; Van~Voorhis,~T. Atom-based bootstrap embedding for molecules.
  \emph{The journal of physical chemistry letters} \textbf{2019}, \emph{10},
  6368--6374\relax
\mciteBstWouldAddEndPuncttrue
\mciteSetBstMidEndSepPunct{\mcitedefaultmidpunct}
{\mcitedefaultendpunct}{\mcitedefaultseppunct}\relax
\EndOfBibitem
\bibitem[Knizia(2013)]{knizia2013intrinsic}
Knizia,~G. Intrinsic atomic orbitals: An unbiased bridge between quantum theory
  and chemical concepts. \emph{Journal of chemical theory and computation}
  \textbf{2013}, \emph{9}, 4834--4843\relax
\mciteBstWouldAddEndPuncttrue
\mciteSetBstMidEndSepPunct{\mcitedefaultmidpunct}
{\mcitedefaultendpunct}{\mcitedefaultseppunct}\relax
\EndOfBibitem
\bibitem[Nusspickel and Booth(2022)Nusspickel, and
  Booth]{nusspickel2022systematic}
Nusspickel,~M.; Booth,~G.~H. Systematic improvability in quantum embedding for
  real materials. \emph{Physical Review X} \textbf{2022}, \emph{12},
  011046\relax
\mciteBstWouldAddEndPuncttrue
\mciteSetBstMidEndSepPunct{\mcitedefaultmidpunct}
{\mcitedefaultendpunct}{\mcitedefaultseppunct}\relax
\EndOfBibitem
\bibitem[Parisen~Toldin and Assaad(2018)Parisen~Toldin, and
  Assaad]{toldin2018entanglement}
Parisen~Toldin,~F.; Assaad,~F.~F. Entanglement Hamiltonian of Interacting
  Fermionic Models. \emph{Physical Review Letters} \textbf{2018}, \emph{121},
  200602\relax
\mciteBstWouldAddEndPuncttrue
\mciteSetBstMidEndSepPunct{\mcitedefaultmidpunct}
{\mcitedefaultendpunct}{\mcitedefaultseppunct}\relax
\EndOfBibitem
\bibitem[Vogiatzis \latin{et~al.}(2017)Vogiatzis, Ma, Olsen, Gagliardi, and
  De~Jong]{vogiatzis2017pushing}
Vogiatzis,~K.~D.; Ma,~D.; Olsen,~J.; Gagliardi,~L.; De~Jong,~W.~A. Pushing
  configuration-interaction to the limit: Towards massively parallel MCSCF
  calculations. \emph{The Journal of Chemical Physics} \textbf{2017},
  \emph{147}, 184111\relax
\mciteBstWouldAddEndPuncttrue
\mciteSetBstMidEndSepPunct{\mcitedefaultmidpunct}
{\mcitedefaultendpunct}{\mcitedefaultseppunct}\relax
\EndOfBibitem
\bibitem[Bartlett and Musia\l{}(2007)Bartlett, and
  Musia\l{}]{bartlett2007coupled}
Bartlett,~R.~J.; Musia\l{},~M. Coupled-cluster theory in quantum chemistry.
  \emph{Review of Modern Physics} \textbf{2007}, \emph{79}, 291--352\relax
\mciteBstWouldAddEndPuncttrue
\mciteSetBstMidEndSepPunct{\mcitedefaultmidpunct}
{\mcitedefaultendpunct}{\mcitedefaultseppunct}\relax
\EndOfBibitem
\bibitem[Morales-Silva \latin{et~al.}(2021)Morales-Silva, Jordan, Shulenburger,
  and Wagner]{morales2021frontiers}
Morales-Silva,~M.~A.; Jordan,~K.~D.; Shulenburger,~L.; Wagner,~L.~K. Frontiers
  of stochastic electronic structure calculations. \emph{The Journal of
  Chemical Physics} \textbf{2021}, \emph{154}, 170401\relax
\mciteBstWouldAddEndPuncttrue
\mciteSetBstMidEndSepPunct{\mcitedefaultmidpunct}
{\mcitedefaultendpunct}{\mcitedefaultseppunct}\relax
\EndOfBibitem
\bibitem[Lee \latin{et~al.}(2022)Lee, Pham, and Reichman]{lee_twenty_2022}
Lee,~J.; Pham,~H.~Q.; Reichman,~D.~R. Twenty {Years} of {Auxiliary}-{Field}
  {Quantum} {Monte} {Carlo} in {Quantum} {Chemistry}: {An} {Overview} and
  {Assessment} on {Main} {Group} {Chemistry} and {Bond}-{Breaking}.
  \emph{Journal of Chemical Theory and Computation} \textbf{2022}, Publisher:
  American Chemical Society\relax
\mciteBstWouldAddEndPuncttrue
\mciteSetBstMidEndSepPunct{\mcitedefaultmidpunct}
{\mcitedefaultendpunct}{\mcitedefaultseppunct}\relax
\EndOfBibitem
\bibitem[Shee \latin{et~al.}(2019)Shee, Rudshteyn, Arthur, Zhang, Reichman, and
  Friesner]{shee2019achieving}
Shee,~J.; Rudshteyn,~B.; Arthur,~E.~J.; Zhang,~S.; Reichman,~D.~R.;
  Friesner,~R.~A. On achieving high accuracy in quantum chemical calculations
  of 3 d transition metal-containing systems: a comparison of auxiliary-field
  quantum monte carlo with coupled cluster, density functional theory, and
  experiment for diatomic molecules. \emph{Journal of Chemical Theory and
  Computation} \textbf{2019}, \emph{15}, 2346--2358\relax
\mciteBstWouldAddEndPuncttrue
\mciteSetBstMidEndSepPunct{\mcitedefaultmidpunct}
{\mcitedefaultendpunct}{\mcitedefaultseppunct}\relax
\EndOfBibitem
\bibitem[Liu \latin{et~al.}(2018)Liu, Cho, and Rubenstein]{liu2018ab}
Liu,~Y.; Cho,~M.; Rubenstein,~B. Ab initio finite temperature auxiliary field
  quantum Monte Carlo. \emph{Journal of Chemical Theory and Computation}
  \textbf{2018}, \emph{14}, 4722--4732\relax
\mciteBstWouldAddEndPuncttrue
\mciteSetBstMidEndSepPunct{\mcitedefaultmidpunct}
{\mcitedefaultendpunct}{\mcitedefaultseppunct}\relax
\EndOfBibitem
\bibitem[Nightingale and Umrigar(1998)Nightingale, and
  Umrigar]{nightingale1998quantum}
Nightingale,~M.~P.; Umrigar,~C.~J. \emph{Quantum Monte Carlo Methods in Physics
  and Chemistry}; Springer Science \& Business Media, 1998\relax
\mciteBstWouldAddEndPuncttrue
\mciteSetBstMidEndSepPunct{\mcitedefaultmidpunct}
{\mcitedefaultendpunct}{\mcitedefaultseppunct}\relax
\EndOfBibitem
\bibitem[Foulkes \latin{et~al.}(2001)Foulkes, Mitas, Needs, and
  Rajagopal]{foulkes2001quantum}
Foulkes,~W. M.~C.; Mitas,~L.; Needs,~R.~J.; Rajagopal,~G. Quantum Monte Carlo
  simulations of solids. \emph{Review of Modern Physics} \textbf{2001},
  \emph{73}, 33--83\relax
\mciteBstWouldAddEndPuncttrue
\mciteSetBstMidEndSepPunct{\mcitedefaultmidpunct}
{\mcitedefaultendpunct}{\mcitedefaultseppunct}\relax
\EndOfBibitem
\bibitem[Huang \latin{et~al.}(2020)Huang, Kueng, and
  Preskill]{huang2020predicting}
Huang,~H.-Y.; Kueng,~R.; Preskill,~J. Predicting many properties of a quantum
  system from very few measurements. \emph{Nature Physics} \textbf{2020},
  \emph{16}, 1050--1057\relax
\mciteBstWouldAddEndPuncttrue
\mciteSetBstMidEndSepPunct{\mcitedefaultmidpunct}
{\mcitedefaultendpunct}{\mcitedefaultseppunct}\relax
\EndOfBibitem
\bibitem[Wagner and Mitas(2007)Wagner, and Mitas]{wagner2007energetics}
Wagner,~L.~K.; Mitas,~L. Energetics and dipole moment of transition metal
  monoxides by quantum Monte Carlo. \emph{The Journal of chemical physics}
  \textbf{2007}, \emph{126}, 034105\relax
\mciteBstWouldAddEndPuncttrue
\mciteSetBstMidEndSepPunct{\mcitedefaultmidpunct}
{\mcitedefaultendpunct}{\mcitedefaultseppunct}\relax
\EndOfBibitem
\bibitem[Wheeler \latin{et~al.}(2021)Wheeler, Pathak, Kleiner, Yuan, Rodrigues,
  Lorsung, Krongchon, Chang, Zhou, Busemeyer, Williams, Muñoz, Chow, and
  Wagner]{wheeler2021pyqmc}
Wheeler,~W.~A.; Pathak,~S.; Kleiner,~K.; Yuan,~S.; Rodrigues,~J. N.~B.;
  Lorsung,~C.; Krongchon,~K.; Chang,~Y.; Zhou,~Y.; Busemeyer,~B.;
  Williams,~K.~T.; Muñoz,~A.; Chow,~C.~Y.; Wagner,~L.~K. PyQMC: an all-Python
  real-space quantum Monte Carlo code, v0.5.1. 2021;
  \url{https://github.com/WagnerGroup/pyqmc}\relax
\mciteBstWouldAddEndPuncttrue
\mciteSetBstMidEndSepPunct{\mcitedefaultmidpunct}
{\mcitedefaultendpunct}{\mcitedefaultseppunct}\relax
\EndOfBibitem
\bibitem[Tranter \latin{et~al.}(2018)Tranter, Love, Mintert, and
  Coveney]{tranter2018comparison}
Tranter,~A.; Love,~P.~J.; Mintert,~F.; Coveney,~P.~V. A comparison of the
  bravyi--kitaev and jordan--wigner transformations for the quantum simulation
  of quantum chemistry. \emph{Journal of Chemical Theory and Computation}
  \textbf{2018}, \emph{14}, 5617--5630\relax
\mciteBstWouldAddEndPuncttrue
\mciteSetBstMidEndSepPunct{\mcitedefaultmidpunct}
{\mcitedefaultendpunct}{\mcitedefaultseppunct}\relax
\EndOfBibitem
\bibitem[Edmiston and Ruedenberg(1963)Edmiston, and
  Ruedenberg]{edmiston1963localized}
Edmiston,~C.; Ruedenberg,~K. Localized Atomic and Molecular Orbitals.
  \emph{Review of Modern Physics} \textbf{1963}, \emph{35}, 457--464\relax
\mciteBstWouldAddEndPuncttrue
\mciteSetBstMidEndSepPunct{\mcitedefaultmidpunct}
{\mcitedefaultendpunct}{\mcitedefaultseppunct}\relax
\EndOfBibitem
\bibitem[Wannier(1962)]{wannier1962dynamics}
Wannier,~G.~H. Dynamics of Band Electrons in Electric and Magnetic Fields.
  \emph{Review of Modern Physics} \textbf{1962}, \emph{34}, 645--655\relax
\mciteBstWouldAddEndPuncttrue
\mciteSetBstMidEndSepPunct{\mcitedefaultmidpunct}
{\mcitedefaultendpunct}{\mcitedefaultseppunct}\relax
\EndOfBibitem
\bibitem[Nielsen and Chuang(2002)Nielsen, and Chuang]{nielsen2002quantum}
Nielsen,~M.~A.; Chuang,~I. \emph{Quantum Computation and Quantum Information};
  Cambridge University Press, 2002\relax
\mciteBstWouldAddEndPuncttrue
\mciteSetBstMidEndSepPunct{\mcitedefaultmidpunct}
{\mcitedefaultendpunct}{\mcitedefaultseppunct}\relax
\EndOfBibitem
\bibitem[Mazziotti(2012)]{mazziotti2012two}
Mazziotti,~D.~A. Two-electron reduced density matrix as the basic variable in
  many-electron quantum chemistry and physics. \emph{Chemical Reviews}
  \textbf{2012}, \emph{112}, 244--262\relax
\mciteBstWouldAddEndPuncttrue
\mciteSetBstMidEndSepPunct{\mcitedefaultmidpunct}
{\mcitedefaultendpunct}{\mcitedefaultseppunct}\relax
\EndOfBibitem
\bibitem[Bertlmann and Krammer(2008)Bertlmann, and Krammer]{bertlmann2008block}
Bertlmann,~R.~A.; Krammer,~P. Bloch vectors for qudits. \emph{Journal of
  Physics A: Mathematical and Theoretical} \textbf{2008}, \emph{41},
  235303\relax
\mciteBstWouldAddEndPuncttrue
\mciteSetBstMidEndSepPunct{\mcitedefaultmidpunct}
{\mcitedefaultendpunct}{\mcitedefaultseppunct}\relax
\EndOfBibitem
\bibitem[Conn \latin{et~al.}(2009)Conn, Scheinberg, and
  Vicente]{conn2009introduction}
Conn,~A.~R.; Scheinberg,~K.; Vicente,~L.~N. \emph{Introduction to
  Derivative-Free Optimization}; SIAM, 2009\relax
\mciteBstWouldAddEndPuncttrue
\mciteSetBstMidEndSepPunct{\mcitedefaultmidpunct}
{\mcitedefaultendpunct}{\mcitedefaultseppunct}\relax
\EndOfBibitem
\bibitem[Fanizza \latin{et~al.}(2020)Fanizza, Rosati, Skotiniotis, Calsamiglia,
  and Giovannetti]{Fanizza2020beyond}
Fanizza,~M.; Rosati,~M.; Skotiniotis,~M.; Calsamiglia,~J.; Giovannetti,~V.
  Beyond the Swap Test: Optimal Estimation of Quantum State Overlap.
  \emph{Physical Review Letters} \textbf{2020}, \emph{124}\relax
\mciteBstWouldAddEndPuncttrue
\mciteSetBstMidEndSepPunct{\mcitedefaultmidpunct}
{\mcitedefaultendpunct}{\mcitedefaultseppunct}\relax
\EndOfBibitem
\bibitem[Harrow and Montanaro(2013)Harrow, and Montanaro]{harrow2013testing}
Harrow,~A.~W.; Montanaro,~A. Testing product states, quantum Merlin-Arthur
  games and tensor optimization. \emph{Journal of the ACM (JACM)}
  \textbf{2013}, \emph{60}, 1--43\relax
\mciteBstWouldAddEndPuncttrue
\mciteSetBstMidEndSepPunct{\mcitedefaultmidpunct}
{\mcitedefaultendpunct}{\mcitedefaultseppunct}\relax
\EndOfBibitem
\bibitem[Mangasarian and Fromovitz(1967)Mangasarian, and
  Fromovitz]{mangasarian1967fritz}
Mangasarian,~O.~L.; Fromovitz,~S. The Fritz John necessary optimality
  conditions in the presence of equality and inequality constraints.
  \emph{Journal of Mathematical Analysis and Applications} \textbf{1967},
  \emph{17}, 37--47\relax
\mciteBstWouldAddEndPuncttrue
\mciteSetBstMidEndSepPunct{\mcitedefaultmidpunct}
{\mcitedefaultendpunct}{\mcitedefaultseppunct}\relax
\EndOfBibitem
\bibitem[Bertsekas(2016)]{bertsekas2016nonlinear}
Bertsekas,~D. \emph{Nonlinear Programming}; Athena scientific optimization and
  computation series; Athena Scientific, 2016; pp 317--330\relax
\mciteBstWouldAddEndPuncttrue
\mciteSetBstMidEndSepPunct{\mcitedefaultmidpunct}
{\mcitedefaultendpunct}{\mcitedefaultseppunct}\relax
\EndOfBibitem
\bibitem[Staszczak \latin{et~al.}(2010)Staszczak, Stoitsov, Baran, and
  Nazarewicz]{staszczak2010augmented}
Staszczak,~A.; Stoitsov,~M.; Baran,~A.; Nazarewicz,~W. Augmented Lagrangian
  method for constrained nuclear density functional theory. \emph{The European
  Physical Journal A} \textbf{2010}, \emph{46}, 85--90\relax
\mciteBstWouldAddEndPuncttrue
\mciteSetBstMidEndSepPunct{\mcitedefaultmidpunct}
{\mcitedefaultendpunct}{\mcitedefaultseppunct}\relax
\EndOfBibitem
\bibitem[Kuroiwa and Nakagawa(2021)Kuroiwa, and Nakagawa]{kuroiwa2021penalty}
Kuroiwa,~K.; Nakagawa,~Y.~O. Penalty methods for a variational quantum
  eigensolver. \emph{Physical Review Research} \textbf{2021}, \emph{3},
  013197\relax
\mciteBstWouldAddEndPuncttrue
\mciteSetBstMidEndSepPunct{\mcitedefaultmidpunct}
{\mcitedefaultendpunct}{\mcitedefaultseppunct}\relax
\EndOfBibitem
\bibitem[Liu \latin{et~al.}(2022)Liu, Dutt, Tao, and Chin]{QBE2022}
Liu,~Y.; Dutt,~A.; Tao,~M.; Chin,~Z.~E. Quantum Bootstrap Embedding.
  \url{https://github.com/yuanliu1/QBootstrapEmbedding}, 2022\relax
\mciteBstWouldAddEndPuncttrue
\mciteSetBstMidEndSepPunct{\mcitedefaultmidpunct}
{\mcitedefaultendpunct}{\mcitedefaultseppunct}\relax
\EndOfBibitem
\bibitem[Byrd \latin{et~al.}(1995)Byrd, Lu, Nocedal, and Zhu]{byrd1995limited}
Byrd,~R.~H.; Lu,~P.; Nocedal,~J.; Zhu,~C. A Limited Memory Algorithm for Bound
  Constrained Optimization. \emph{SIAM Journal on Scientific Computing}
  \textbf{1995}, \emph{16}, 1190--1208\relax
\mciteBstWouldAddEndPuncttrue
\mciteSetBstMidEndSepPunct{\mcitedefaultmidpunct}
{\mcitedefaultendpunct}{\mcitedefaultseppunct}\relax
\EndOfBibitem
\bibitem[Nelder and Mead(1965)Nelder, and Mead]{nelder1965simplex}
Nelder,~J.~A.; Mead,~R. {A Simplex Method for Function Minimization}. \emph{The
  Computer Journal} \textbf{1965}, \emph{7}, 308--313\relax
\mciteBstWouldAddEndPuncttrue
\mciteSetBstMidEndSepPunct{\mcitedefaultmidpunct}
{\mcitedefaultendpunct}{\mcitedefaultseppunct}\relax
\EndOfBibitem
\bibitem[Svore \latin{et~al.}(2013)Svore, Hastings, and
  Freedman]{svore2013faster}
Svore,~K.~M.; Hastings,~M.~B.; Freedman,~M. Faster Phase Estimation.
  \emph{Quantum Information and Computation} \textbf{2013}, \relax
\mciteBstWouldAddEndPunctfalse
\mciteSetBstMidEndSepPunct{\mcitedefaultmidpunct}
{}{\mcitedefaultseppunct}\relax
\EndOfBibitem
\bibitem[Huggins \latin{et~al.}(2022)Huggins, O’Gorman, Rubin, Reichman,
  Babbush, and Lee]{huggins2022unbiasing}
Huggins,~W.~J.; O’Gorman,~B.~A.; Rubin,~N.~C.; Reichman,~D.~R.; Babbush,~R.;
  Lee,~J. Unbiasing fermionic quantum Monte Carlo with a quantum computer.
  \emph{Nature} \textbf{2022}, \emph{603}, 416--420\relax
\mciteBstWouldAddEndPuncttrue
\mciteSetBstMidEndSepPunct{\mcitedefaultmidpunct}
{\mcitedefaultendpunct}{\mcitedefaultseppunct}\relax
\EndOfBibitem
\bibitem[Wocjan and Abeyesinghe(2008)Wocjan, and
  Abeyesinghe]{wocjan2008speedup}
Wocjan,~P.; Abeyesinghe,~A. Speedup via quantum sampling. \emph{Phys. Rev. A}
  \textbf{2008}, \emph{78}, 042336\relax
\mciteBstWouldAddEndPuncttrue
\mciteSetBstMidEndSepPunct{\mcitedefaultmidpunct}
{\mcitedefaultendpunct}{\mcitedefaultseppunct}\relax
\EndOfBibitem
\bibitem[Yung and Aspuru-Guzik(2012)Yung, and Aspuru-Guzik]{yung2012quantum}
Yung,~M.-H.; Aspuru-Guzik,~A. A quantum--quantum Metropolis algorithm.
  \emph{Proceedings of the National Academy of Sciences} \textbf{2012},
  \emph{109}, 754--759\relax
\mciteBstWouldAddEndPuncttrue
\mciteSetBstMidEndSepPunct{\mcitedefaultmidpunct}
{\mcitedefaultendpunct}{\mcitedefaultseppunct}\relax
\EndOfBibitem
\bibitem[Montanaro(2015)]{montanaro2015quantum}
Montanaro,~A. Quantum speedup of Monte Carlo methods. \emph{Proceedings of the
  Royal Society A: Mathematical, Physical and Engineering Sciences}
  \textbf{2015}, \emph{471}, 20150301\relax
\mciteBstWouldAddEndPuncttrue
\mciteSetBstMidEndSepPunct{\mcitedefaultmidpunct}
{\mcitedefaultendpunct}{\mcitedefaultseppunct}\relax
\EndOfBibitem
\bibitem[Yoder \latin{et~al.}(2014)Yoder, Low, and Chuang]{yoder2014fixed}
Yoder,~T.~J.; Low,~G.~H.; Chuang,~I.~L. Fixed-Point Quantum Search with an
  Optimal Number of Queries. \emph{Physical Review Letters} \textbf{2014},
  \emph{113}, 210501\relax
\mciteBstWouldAddEndPuncttrue
\mciteSetBstMidEndSepPunct{\mcitedefaultmidpunct}
{\mcitedefaultendpunct}{\mcitedefaultseppunct}\relax
\EndOfBibitem
\bibitem[Berry \latin{et~al.}(2014)Berry, Childs, Cleve, Kothari, and
  Somma]{berry2014exponential}
Berry,~D.~W.; Childs,~A.~M.; Cleve,~R.; Kothari,~R.; Somma,~R.~D. Exponential
  improvement in precision for simulating sparse Hamiltonians. Proceedings of
  the forty-sixth annual ACM symposium on Theory of computing. 2014; pp
  283--292\relax
\mciteBstWouldAddEndPuncttrue
\mciteSetBstMidEndSepPunct{\mcitedefaultmidpunct}
{\mcitedefaultendpunct}{\mcitedefaultseppunct}\relax
\EndOfBibitem
\bibitem[Qi and Ranard(2021)Qi, and Ranard]{qi2021emergent}
Qi,~X.-L.; Ranard,~D. Emergent classicality in general multipartite states and
  channels. \emph{Quantum} \textbf{2021}, \emph{5}, 555\relax
\mciteBstWouldAddEndPuncttrue
\mciteSetBstMidEndSepPunct{\mcitedefaultmidpunct}
{\mcitedefaultendpunct}{\mcitedefaultseppunct}\relax
\EndOfBibitem
\bibitem[Nusspickel \latin{et~al.}(2022)Nusspickel, Ibrahim, and
  Booth]{nusspickel2022effective}
Nusspickel,~M.; Ibrahim,~B.; Booth,~G.~H. On the effective reconstruction of
  expectation values from ab initio quantum embedding. \emph{arXiv preprint
  arXiv:2210.14561} \textbf{2022}, \relax
\mciteBstWouldAddEndPunctfalse
\mciteSetBstMidEndSepPunct{\mcitedefaultmidpunct}
{}{\mcitedefaultseppunct}\relax
\EndOfBibitem
\bibitem[Mitra \latin{et~al.}(2021)Mitra, Pham, Pandharkar, Hermes, and
  Gagliardi]{mitra2021excited}
Mitra,~A.; Pham,~H.~Q.; Pandharkar,~R.; Hermes,~M.~R.; Gagliardi,~L. Excited
  states of crystalline point defects with multireference density matrix
  embedding theory. \emph{The Journal of Physical Chemistry Letters}
  \textbf{2021}, \emph{12}, 11688--11694\relax
\mciteBstWouldAddEndPuncttrue
\mciteSetBstMidEndSepPunct{\mcitedefaultmidpunct}
{\mcitedefaultendpunct}{\mcitedefaultseppunct}\relax
\EndOfBibitem
\bibitem[Zhang(1999)]{zhang1999finite}
Zhang,~S. Finite-temperature Monte Carlo calculations for systems with
  fermions. \emph{Physical Review Letters} \textbf{1999}, \emph{83}, 2777\relax
\mciteBstWouldAddEndPuncttrue
\mciteSetBstMidEndSepPunct{\mcitedefaultmidpunct}
{\mcitedefaultendpunct}{\mcitedefaultseppunct}\relax
\EndOfBibitem
\bibitem[Sun \latin{et~al.}(2020)Sun, Ray, Cui, Stoudenmire, Ferrero, and
  Chan]{sun2020finite}
Sun,~C.; Ray,~U.; Cui,~Z.-H.; Stoudenmire,~M.; Ferrero,~M.; Chan,~G. K.-L.
  Finite-temperature density matrix embedding theory. \emph{Physical Review B}
  \textbf{2020}, \emph{101}, 075131\relax
\mciteBstWouldAddEndPuncttrue
\mciteSetBstMidEndSepPunct{\mcitedefaultmidpunct}
{\mcitedefaultendpunct}{\mcitedefaultseppunct}\relax
\EndOfBibitem
\bibitem[Freeze \latin{et~al.}(2019)Freeze, Kelly, and
  Batista]{freeze2019search}
Freeze,~J.~G.; Kelly,~H.~R.; Batista,~V.~S. Search for catalysts by inverse
  design: artificial intelligence, mountain climbers, and alchemists.
  \emph{Chemical Reviews} \textbf{2019}, \emph{119}, 6595--6612\relax
\mciteBstWouldAddEndPuncttrue
\mciteSetBstMidEndSepPunct{\mcitedefaultmidpunct}
{\mcitedefaultendpunct}{\mcitedefaultseppunct}\relax
\EndOfBibitem
\bibitem[Zhou \latin{et~al.}(2012)Zhou, Long, and Yaghi]{zhou2012introduction}
Zhou,~H.-C.; Long,~J.~R.; Yaghi,~O.~M. Introduction to Metal--Organic
  Frameworks. 2012; \url{https://pubs.acs.org/doi/10.1021/cr300014x}\relax
\mciteBstWouldAddEndPuncttrue
\mciteSetBstMidEndSepPunct{\mcitedefaultmidpunct}
{\mcitedefaultendpunct}{\mcitedefaultseppunct}\relax
\EndOfBibitem
\bibitem[Warshel(2014)]{warshel2014multiscale}
Warshel,~A. Multiscale modeling of biological functions: from enzymes to
  molecular machines (Nobel Lecture). \emph{Angewandte Chemie International
  Edition} \textbf{2014}, \emph{53}, 10020--10031\relax
\mciteBstWouldAddEndPuncttrue
\mciteSetBstMidEndSepPunct{\mcitedefaultmidpunct}
{\mcitedefaultendpunct}{\mcitedefaultseppunct}\relax
\EndOfBibitem
\bibitem[Proppe \latin{et~al.}(2020)Proppe, Li, Aspuru-Guzik, Berlinguette,
  Chang, Cogdell, Doyle, Flick, Gabor, van Grondelle, Hammes-Schiffer, Jaffer,
  Kelley, Leclerc, Leo, Mallouk, Narang, Schlau-Cohen, Scholes, Vojvodic, Yam,
  Yang, and Sargent]{proppe2020bioinspiration}
Proppe,~A.~H.; Li,~Y.~C.; Aspuru-Guzik,~A.; Berlinguette,~C.~P.; Chang,~C.~J.;
  Cogdell,~R.; Doyle,~A.~G.; Flick,~J.; Gabor,~N.~M.; van Grondelle,~R.;
  Hammes-Schiffer,~S.; Jaffer,~S.~A.; Kelley,~S.~O.; Leclerc,~M.; Leo,~K.;
  Mallouk,~T.~E.; Narang,~P.; Schlau-Cohen,~G.~S.; Scholes,~G.~D.;
  Vojvodic,~A.; Yam,~V. W.-W.; Yang,~J.~Y.; Sargent,~E.~H. Bioinspiration in
  light harvesting and catalysis. \emph{Nature Reviews Materials}
  \textbf{2020}, \emph{5}, 828--846\relax
\mciteBstWouldAddEndPuncttrue
\mciteSetBstMidEndSepPunct{\mcitedefaultmidpunct}
{\mcitedefaultendpunct}{\mcitedefaultseppunct}\relax
\EndOfBibitem
\bibitem[Pham \latin{et~al.}(2019)Pham, Hermes, and
  Gagliardi]{pham2019periodic}
Pham,~H.~Q.; Hermes,~M.~R.; Gagliardi,~L. Periodic electronic structure
  calculations with the density matrix embedding theory. \emph{Journal of
  Chemical Theory and Computation} \textbf{2019}, \emph{16}, 130--140\relax
\mciteBstWouldAddEndPuncttrue
\mciteSetBstMidEndSepPunct{\mcitedefaultmidpunct}
{\mcitedefaultendpunct}{\mcitedefaultseppunct}\relax
\EndOfBibitem
\bibitem[Chibani \latin{et~al.}(2016)Chibani, Ren, Scheffler, and
  Rinke]{chibani2016self}
Chibani,~W.; Ren,~X.; Scheffler,~M.; Rinke,~P. Self-consistent Green's function
  embedding for advanced electronic structure methods based on a dynamical
  mean-field concept. \emph{Physical Review B} \textbf{2016}, \emph{93},
  165106\relax
\mciteBstWouldAddEndPuncttrue
\mciteSetBstMidEndSepPunct{\mcitedefaultmidpunct}
{\mcitedefaultendpunct}{\mcitedefaultseppunct}\relax
\EndOfBibitem
\bibitem[Head-Marsden \latin{et~al.}(2020)Head-Marsden, Flick, Ciccarino, and
  Narang]{head2020quantum}
Head-Marsden,~K.; Flick,~J.; Ciccarino,~C.~J.; Narang,~P. Quantum information
  and algorithms for correlated quantum matter. \emph{Chemical Reviews}
  \textbf{2020}, \emph{121}, 3061--3120\relax
\mciteBstWouldAddEndPuncttrue
\mciteSetBstMidEndSepPunct{\mcitedefaultmidpunct}
{\mcitedefaultendpunct}{\mcitedefaultseppunct}\relax
\EndOfBibitem
\bibitem[Qin \latin{et~al.}(2022)Qin, Sch{\"a}fer, Andergassen, Corboz, and
  Gull]{qin2022hubbard}
Qin,~M.; Sch{\"a}fer,~T.; Andergassen,~S.; Corboz,~P.; Gull,~E. The Hubbard
  model: A computational perspective. \emph{Annual Review of Condensed Matter
  Physics} \textbf{2022}, \emph{13}, 275--302\relax
\mciteBstWouldAddEndPuncttrue
\mciteSetBstMidEndSepPunct{\mcitedefaultmidpunct}
{\mcitedefaultendpunct}{\mcitedefaultseppunct}\relax
\EndOfBibitem
\bibitem[Ding \latin{et~al.}(2020)Ding, Mardazad, Das, Szalay, Schollwöck,
  Zimbor{\'a}s, and Schilling]{ding2020concept}
Ding,~L.; Mardazad,~S.; Das,~S.; Szalay,~S.; Schollwöck,~U.; Zimbor{\'a}s,~Z.;
  Schilling,~C. Concept of orbital entanglement and correlation in quantum
  chemistry. \emph{Journal of Chemical Theory and Computation} \textbf{2020},
  \emph{17}, 79--95\relax
\mciteBstWouldAddEndPuncttrue
\mciteSetBstMidEndSepPunct{\mcitedefaultmidpunct}
{\mcitedefaultendpunct}{\mcitedefaultseppunct}\relax
\EndOfBibitem
\bibitem[Ding and Schilling(2020)Ding, and Schilling]{ding2020correlation}
Ding,~L.; Schilling,~C. Correlation paradox of the dissociation limit: A
  quantum information perspective. \emph{Journal of Chemical Theory and
  Computation} \textbf{2020}, \emph{16}, 4159--4175\relax
\mciteBstWouldAddEndPuncttrue
\mciteSetBstMidEndSepPunct{\mcitedefaultmidpunct}
{\mcitedefaultendpunct}{\mcitedefaultseppunct}\relax
\EndOfBibitem
\bibitem[Wilde(2013)]{wilde2013quantum}
Wilde,~M.~M. \emph{Quantum Information Theory}; Cambridge University Press,
  2013\relax
\mciteBstWouldAddEndPuncttrue
\mciteSetBstMidEndSepPunct{\mcitedefaultmidpunct}
{\mcitedefaultendpunct}{\mcitedefaultseppunct}\relax
\EndOfBibitem
\bibitem[Harrow(2020)]{harrow2020small}
Harrow,~A.~W. Small quantum computers and large classical data sets.
  \emph{arXiv preprint arXiv:2004.00026} \textbf{2020}, \relax
\mciteBstWouldAddEndPunctfalse
\mciteSetBstMidEndSepPunct{\mcitedefaultmidpunct}
{}{\mcitedefaultseppunct}\relax
\EndOfBibitem
\bibitem[Song \latin{et~al.}(2021)Song, Wie{\'s}niak, Liu, Paw{\l}owski, Lee,
  Kim, and Bang]{song2021tangible}
Song,~W.; Wie{\'s}niak,~M.; Liu,~N.; Paw{\l}owski,~M.; Lee,~J.; Kim,~J.;
  Bang,~J. Tangible reduction in learning sample complexity with large
  classical samples and small quantum system. \emph{Quantum Information
  Processing} \textbf{2021}, \emph{20}, 275\relax
\mciteBstWouldAddEndPuncttrue
\mciteSetBstMidEndSepPunct{\mcitedefaultmidpunct}
{\mcitedefaultendpunct}{\mcitedefaultseppunct}\relax
\EndOfBibitem
\end{mcitethebibliography}


\makeatletter\@input{arxiv-help-si.tex}\makeatother
\end{document}